\documentclass[11pt]{article}

\usepackage{fullpage,url}
\newenvironment{descit}[1]{\begin{quote} \textit{#1}}{\end{quote}}

\ifx\undefined\psfig\else \fi

\edef\psfigRestoreAt{\catcode`@=\number\catcode`@\relax}
\catcode`\@=11\relax
\newwrite\@unused
\def\typeout#1{{\let\protect\string\immediate\write\@unused{#1}}}
\def\figurepath{./}

\def\@nnil{\@nil}
\def\@empty{}
\def\@psdonoop#1\@@#2#3{}
\def\@psdo#1:=#2\do#3{\edef\@psdotmp{#2}\ifx\@psdotmp\@empty \else
    \expandafter\@psdoloop#2,\@nil,\@nil\@@#1{#3}\fi}
\def\@psdoloop#1,#2,#3\@@#4#5{\def#4{#1}\ifx #4\@nnil \else
       #5\def#4{#2}\ifx #4\@nnil \else#5\@ipsdoloop #3\@@#4{#5}\fi\fi}
\def\@ipsdoloop#1,#2\@@#3#4{\def#3{#1}\ifx #3\@nnil 
       \let\@nextwhile=\@psdonoop \else
      #4\relax\let\@nextwhile=\@ipsdoloop\fi\@nextwhile#2\@@#3{#4}}
\def\@tpsdo#1:=#2\do#3{\xdef\@psdotmp{#2}\ifx\@psdotmp\@empty \else
    \@tpsdoloop#2\@nil\@nil\@@#1{#3}\fi}
\def\@tpsdoloop#1#2\@@#3#4{\def#3{#1}\ifx #3\@nnil 
       \let\@nextwhile=\@psdonoop \else
      #4\relax\let\@nextwhile=\@tpsdoloop\fi\@nextwhile#2\@@#3{#4}}
\newread\ps@stream
\newif\ifnot@eof       % continue looking for the bounding box?
\newif\if@noisy        % report what you're making?
\newif\if@atend        % %%BoundingBox: has (at end) specification
\newif\if@psfile       % does this look like a PostScript file?
{\catcode`\%=12\global\gdef\epsf@start{%!}}
\def\epsf@PS{PS}
\def\epsf@getbb#1{%
\openin\ps@stream=#1
\ifeof\ps@stream\typeout{Error, File #1 not found}\else
   {\not@eoftrue \chardef\other=12
    \def\do##1{\catcode`##1=\other}\dospecials \catcode`\ =10
    \loop
       \if@psfile
	  \read\ps@stream to \epsf@fileline
       \else{
	  \obeyspaces
          \read\ps@stream to \epsf@tmp\global\let\epsf@fileline\epsf@tmp}
       \fi
       \ifeof\ps@stream\not@eoffalse\else
       \if@psfile\else
       \expandafter\epsf@test\epsf@fileline:. \\%
       \fi
          \expandafter\epsf@aux\epsf@fileline:. \\%
       \fi
   \ifnot@eof\repeat
   }\closein\ps@stream\fi}%
\long\def\epsf@test#1#2#3:#4\\{\def\epsf@testit{#1#2}
			\ifx\epsf@testit\epsf@start\else
\typeout{Warning! File does not start with `\epsf@start'.  It may not be a PostScript file.}
			\fi
			\@psfiletrue} % don't test after 1st line
{\catcode`\%=12\global\let\epsf@percent=%\global\def\epsf@bblit{%BoundingBox}}
\long\def\epsf@aux#1#2:#3\\{\ifx#1\epsf@percent
   \def\epsf@testit{#2}\ifx\epsf@testit\epsf@bblit
	\@atendfalse
        \epsf@atend #3 . \\%
	\if@atend	
	   \if@verbose{
		\typeout{psfig: found `(atend)'; continuing search}
	   }\fi
        \else
        \epsf@grab #3 . . . \\%
        \not@eoffalse
        \global\no@bbfalse
        \fi
   \fi\fi}%
\def\epsf@grab #1 #2 #3 #4 #5\\{%
   \global\def\epsf@llx{#1}\ifx\epsf@llx\empty
      \epsf@grab #2 #3 #4 #5 .\\\else
   \global\def\epsf@lly{#2}%
   \global\def\epsf@urx{#3}\global\def\epsf@ury{#4}\fi}%
\def\epsf@atendlit{(atend)} 
\def\epsf@atend #1 #2 #3\\{%
   \def\epsf@tmp{#1}\ifx\epsf@tmp\empty
      \epsf@atend #2 #3 .\\\else
   \ifx\epsf@tmp\epsf@atendlit\@atendtrue\fi\fi}

\chardef\letter = 11
\chardef\other = 12

\newif \ifdebug %%% turn me on to see TeX hard at work ...
\newif\ifc@mpute %%% don't need to compute some values
\c@mputetrue % but assume that we do

\let\then = \relax
\def\r@dian{pt }
\let\r@dians = \r@dian
\let\dimensionless@nit = \r@dian
\let\dimensionless@nits = \dimensionless@nit
\def\internal@nit{sp }
\let\internal@nits = \internal@nit
\newif\ifstillc@nverging
\def \Mess@ge #1{\ifdebug \then \message {#1} \fi}

{ %%% Things that need abnormal catcodes %%%
	\catcode `\@ = \letter
	\gdef \nodimen {\expandafter \n@dimen \the \dimen}
	\gdef \term #1 #2 #3%
	       {\edef \t@ {\the #1}%%% freeze parameter 1 (count, by value)
		\edef \t@@ {\expandafter \n@dimen \the #2\r@dian}%
		\t@rm {\t@} {\t@@} {#3}%
	       }
	\gdef \t@rm #1 #2 #3%
	       {{%
		\count 0 = 0
		\dimen 0 = 1 \dimensionless@nit
		\dimen 2 = #2\relax
		\Mess@ge {Calculating term #1 of \nodimen 2}%
		\loop
		\ifnum	\count 0 < #1
		\then	\advance \count 0 by 1
			\Mess@ge {Iteration \the \count 0 \space}%
			\Multiply \dimen 0 by {\dimen 2}%
			\Mess@ge {After multiplication, term = \nodimen 0}%
			\Divide \dimen 0 by {\count 0}%
			\Mess@ge {After division, term = \nodimen 0}%
		\repeat
		\Mess@ge {Final value for term #1 of 
				\nodimen 2 \space is \nodimen 0}%
		\xdef \Term {#3 = \nodimen 0 \r@dians}%
		\aftergroup \Term
	       }}
	\catcode `\p = \other
	\catcode `\t = \other
	\gdef \n@dimen #1pt{#1} %%% throw away the ``pt''
}

\def \Divide #1by #2{\divide #1 by #2} %%% just a synonym

\def \Multiply #1by #2%%% allows division of a dimen by a dimen
       {{%%% should really freeze parameter 2 (dimen, passed by value)
	\count 0 = #1\relax
	\count 2 = #2\relax
	\count 4 = 65536
	\Mess@ge {Before scaling, count 0 = \the \count 0 \space and
			count 2 = \the \count 2}%
	\ifnum	\count 0 > 32767 %%% do our best to avoid overflow
	\then	\divide \count 0 by 4
		\divide \count 4 by 4
	\else	\ifnum	\count 0 < -32767
		\then	\divide \count 0 by 4
			\divide \count 4 by 4
		\else
		\fi
	\fi
	\ifnum	\count 2 > 32767 %%% while retaining reasonable accuracy
	\then	\divide \count 2 by 4
		\divide \count 4 by 4
	\else	\ifnum	\count 2 < -32767
		\then	\divide \count 2 by 4
			\divide \count 4 by 4
		\else
		\fi
	\fi
	\multiply \count 0 by \count 2
	\divide \count 0 by \count 4
	\xdef \product {#1 = \the \count 0 \internal@nits}%
	\aftergroup \product
       }}

\def\r@duce{\ifdim\dimen0 > 90\r@dian \then   % sin(x) = sin(180-x)
		\multiply\dimen0 by -1
		\advance\dimen0 by 180\r@dian
		\r@duce
	    \else \ifdim\dimen0 < -90\r@dian \then  % sin(x) = sin(360+x)
		\advance\dimen0 by 360\r@dian
		\r@duce
		\fi
	    \fi}

\def\Sine#1%
       {{%
	\dimen 0 = #1 \r@dian
	\r@duce
	\ifdim\dimen0 = -90\r@dian \then
	   \dimen4 = -1\r@dian
	   \c@mputefalse
	\fi
	\ifdim\dimen0 = 90\r@dian \then
	   \dimen4 = 1\r@dian
	   \c@mputefalse
	\fi
	\ifdim\dimen0 = 0\r@dian \then
	   \dimen4 = 0\r@dian
	   \c@mputefalse
	\fi
	\ifc@mpute \then
		\divide\dimen0 by 180
		\dimen0=3.141592654\dimen0
		\dimen 2 = 3.1415926535897963\r@dian %%% a well-known constant
		\divide\dimen 2 by 2 %%% we only deal with -pi/2 : pi/2
		\Mess@ge {Sin: calculating Sin of \nodimen 0}%
		\count 0 = 1 %%% see power-series expansion for sine
		\dimen 2 = 1 \r@dian %%% ditto
		\dimen 4 = 0 \r@dian %%% ditto
		\loop
			\ifnum	\dimen 2 = 0 %%% then we've done
			\then	\stillc@nvergingfalse 
			\else	\stillc@nvergingtrue
			\fi
			\ifstillc@nverging %%% then calculate next term
			\then	\term {\count 0} {\dimen 0} {\dimen 2}%
				\advance \count 0 by 2
				\count 2 = \count 0
				\divide \count 2 by 2
				\ifodd	\count 2 %%% signs alternate
				\then	\advance \dimen 4 by \dimen 2
				\else	\advance \dimen 4 by -\dimen 2
				\fi
		\repeat
	\fi		
			\xdef \sine {\nodimen 4}%
       }}

\def\Cosine#1{\ifx\sine\UnDefined\edef\Savesine{\relax}\else
		             \edef\Savesine{\sine}\fi
	{\dimen0=#1\r@dian\multiply\dimen0 by -1
	 \advance\dimen0 by 90\r@dian
	 \Sine{\nodimen 0}
	 \xdef\cosine{\sine}
	 \xdef\sine{\Savesine}}}	      
\def\psdraft{
	\def\@psdraft{0}
}
\def\psfull{
	\def\@psdraft{100}
}

\psfull

\newif\if@draftbox
\def\psnodraftbox{
	\@draftboxfalse
}
\@draftboxtrue

\newif\if@prologfile
\newif\if@postlogfile
\def\pssilent{
	\@noisyfalse
}
\def\psnoisy{
	\@noisytrue
}
\psnoisy
\newif\if@bbllx
\newif\if@bblly
\newif\if@bburx
\newif\if@bbury
\newif\if@height
\newif\if@width
\newif\if@rheight
\newif\if@rwidth
\newif\if@angle
\newif\if@clip
\newif\if@verbose
\newif\if@scale
\def\@p@@sclip#1{\@cliptrue}

\def\@p@@sfile#1{\def\@p@sfile{null}%
	        \openin1=#1
		\ifeof1\closein1%
		       \openin1=\figurepath#1
			\ifeof1\typeout{Error, File #1 not found}
			   \if@bbllx\if@bblly\if@bburx\if@bbury% added 6/91 Rob Russell
			      \def\@p@sfile{#1}%
			   \fi\fi\fi\fi
			\else\closein1
			    \edef\@p@sfile{\figurepath#1}%
                        \fi%
		 \else\closein1%
		       \def\@p@sfile{#1}%
		 \fi}
\def\@p@@sfigure#1{\def\@p@sfile{null}%
	        \openin1=#1
		\ifeof1\closein1%
		       \openin1=\figurepath#1
			\ifeof1\typeout{Error, File #1 not found}
			   \if@bbllx\if@bblly\if@bburx\if@bbury% added 6/91 Rob Russell
			      \def\@p@sfile{#1}%
			   \fi\fi\fi\fi
			\else\closein1
			    \def\@p@sfile{\figurepath#1}%
                        \fi%
		 \else\closein1%
		       \def\@p@sfile{#1}%
		 \fi}

\def\@p@@sbbllx#1{
		\@bbllxtrue
		\dimen100=#1
		\edef\@p@sbbllx{\number\dimen100}
}
\def\@p@@sbblly#1{
		\@bbllytrue
		\dimen100=#1
		\edef\@p@sbblly{\number\dimen100}
}
\def\@p@@sbburx#1{
		\@bburxtrue
		\dimen100=#1
		\edef\@p@sbburx{\number\dimen100}
}
\def\@p@@sbbury#1{
		\@bburytrue
		\dimen100=#1
		\edef\@p@sbbury{\number\dimen100}
}
\def\@p@@sheight#1{
		\@heighttrue
		\dimen100=#1
   		\edef\@p@sheight{\number\dimen100}
}
\def\@p@@swidth#1{
		\@widthtrue
		\dimen100=#1
		\edef\@p@swidth{\number\dimen100}
}
\def\@p@@srheight#1{
		\@rheighttrue
		\dimen100=#1
		\edef\@p@srheight{\number\dimen100}
}
\def\@p@@srwidth#1{
		\@rwidthtrue
		\dimen100=#1
		\edef\@p@srwidth{\number\dimen100}
}
\def\@p@@sangle#1{
		\@angletrue
		\edef\@p@sangle{#1} %\number\dimen100}
}
\def\@p@@ssilent#1{ 
		\@verbosefalse
}
\def\@p@@sscale#1{
		\def\@p@scale{#1}
		\@scaletrue
}
\def\@p@@sprolog#1{\@prologfiletrue\def\@prologfileval{#1}}
\def\@p@@spostlog#1{\@postlogfiletrue\def\@postlogfileval{#1}}
\def\@cs@name#1{\csname #1\endcsname}
\def\@setparms#1=#2,{\@cs@name{@p@@s#1}{#2}}
\def\ps@init@parms{
		\@bbllxfalse \@bbllyfalse
		\@bburxfalse \@bburyfalse
		\@heightfalse \@widthfalse
		\@rheightfalse \@rwidthfalse
		\@scalefalse
		\def\@p@sbbllx{}\def\@p@sbblly{}
		\def\@p@sbburx{}\def\@p@sbbury{}
		\def\@p@sheight{}\def\@p@swidth{}
		\def\@p@srheight{}\def\@p@srwidth{}
		\def\@p@sangle{0}
		\def\@p@sfile{}
		\def\@p@scost{10}
		\def\@sc{}
		\@prologfilefalse
		\@postlogfilefalse
		\@clipfalse
		\if@noisy
			\@verbosetrue
		\else
			\@verbosefalse
		\fi
}
\def\parse@ps@parms#1{
	 	\@psdo\@psfiga:=#1\do
		   {\expandafter\@setparms\@psfiga,}}
\newif\ifno@bb
\def\bb@missing{
	\if@verbose{
		\typeout{psfig: searching \@p@sfile \space  for bounding box}
	}\fi
	\no@bbtrue
	\epsf@getbb{\@p@sfile}
        \ifno@bb \else \bb@cull\epsf@llx\epsf@lly\epsf@urx\epsf@ury\fi
}	
\def\bb@cull#1#2#3#4{
	\dimen100=#1 bp\edef\@p@sbbllx{\number\dimen100}
	\dimen100=#2 bp\edef\@p@sbblly{\number\dimen100}
	\dimen100=#3 bp\edef\@p@sbburx{\number\dimen100}
	\dimen100=#4 bp\edef\@p@sbbury{\number\dimen100}
	\no@bbfalse
}

\newdimen\p@intvaluex
\newdimen\p@intvaluey
\newdimen\@ffsetvalue
\newdimen\x@ffsetvalue
\newdimen\y@ffsetvalue

\def\compute@offset#1#2{{\dimen0=#1 sp\dimen1=#2 sp
			\advance\dimen1 by -\dimen0
			\dimen1=\sine\dimen1
			\dimen0=\cosine\dimen1
			\ifdim\dimen0<0sp \dimen1=0sp \fi
			\global\@ffsetvalue=\dimen1}}

\def\rotate@#1#2{{\dimen0=#1 sp\dimen1=#2 sp
		  \global\p@intvaluex=\cosine\dimen0
		  \dimen3=\sine\dimen1
		  \global\advance\p@intvaluex by -\dimen3
		  \global\p@intvaluey=\sine\dimen0
		  \dimen3=\cosine\dimen1
		  \global\advance\p@intvaluey by \dimen3
		  }}
\def\compute@bb{
		\no@bbfalse
		\if@bbllx \else \no@bbtrue \fi
		\if@bblly \else \no@bbtrue \fi
		\if@bburx \else \no@bbtrue \fi
		\if@bbury \else \no@bbtrue \fi
		\ifno@bb \bb@missing \fi
		\ifno@bb \typeout{FATAL ERROR: no bb supplied or found}
			\no-bb-error
		\fi
		\if@angle 
			\Sine{\@p@sangle}\Cosine{\@p@sangle}
			\compute@offset{\@p@sbblly}{\@p@sbbury}
			\x@ffsetvalue=\@ffsetvalue
			\compute@offset{\@p@sbburx}{\@p@sbbllx}
			\y@ffsetvalue=\@ffsetvalue

			\rotate@{\@p@sbbllx}{\@p@sbblly}
			\advance\p@intvaluex by -\x@ffsetvalue
			\advance\p@intvaluey by -\y@ffsetvalue
			\edef\@p@sbbllx{\number\p@intvaluex}
			\edef\@p@sbblly{\number\p@intvaluey}

			\rotate@{\@p@sbburx}{\@p@sbbury}
			\advance\p@intvaluex by \x@ffsetvalue
			\advance\p@intvaluey by \y@ffsetvalue
			\edef\@p@sbburx{\number\p@intvaluex}
			\edef\@p@sbbury{\number\p@intvaluey}
			{
			 \count0=\@p@sbbllx \count1=\@p@sbblly
		 	 \count2=\@p@sbburx \count3=\@p@sbbury
			 \dimen0=\@p@sbbllx sp\dimen1=\@p@sbblly sp
		 	 \dimen2=\@p@sbburx sp\dimen3=\@p@sbbury sp
			 \dimen203=\dimen2 \advance\dimen203 by -\dimen0
			 \dimen204=\dimen3 \advance\dimen204 by -\dimen1
			 \ifdim\dimen203<0sp 
			      \count203=\count2 \count2=\count0 
			      \count0=\count203 
			      \global\edef\@p@sbbllx{\number\count0}
			      \global\edef\@p@sbburx{\number\count2}
			 \fi
			 \ifdim\dimen204<0sp 
			       \count204=\count3
			       \count3=\count1
			       \count1=\count204
			       \global\edef\@p@sbblly{\number\count1}
			       \global\edef\@p@sbbury{\number\count3}
			 \fi
			}
		\fi
		\count203=\@p@sbburx
		\count204=\@p@sbbury
		\advance\count203 by -\@p@sbbllx
		\advance\count204 by -\@p@sbblly
		\edef\@bbw{\number\count203}
		\edef\@bbh{\number\count204}
}
\def\in@hundreds#1#2#3{\count240=#2 \count241=#3
		     \count100=\count240	% 100 is first digit #2/#3
		     \divide\count100 by \count241
		     \count101=\count100
		     \multiply\count101 by \count241
		     \advance\count240 by -\count101
		     \multiply\count240 by 10
		     \count101=\count240	%101 is second digit of #2/#3
		     \divide\count101 by \count241
		     \count102=\count101
		     \multiply\count102 by \count241
		     \advance\count240 by -\count102
		     \multiply\count240 by 10
		     \count102=\count240	% 102 is the third digit
		     \divide\count102 by \count241
		     \count200=#1\count205=0
		     \count201=\count200
			\multiply\count201 by \count100
		 	\advance\count205 by \count201
		     \count201=\count200
			\divide\count201 by 10
			\multiply\count201 by \count101
			\advance\count205 by \count201
		     \count201=\count200
			\divide\count201 by 100
			\multiply\count201 by \count102
			\advance\count205 by \count201
		     \edef\@result{\number\count205}
}
\def\@ScaleInHundreds#1{
		\in@hundreds{#1}{\@p@scale}{100}
		\edef#1{\@result}
}
\def\compute@wfromh{
		\in@hundreds{\@p@sheight}{\@bbw}{\@bbh}
		\edef\@p@swidth{\@result}
}
\def\compute@hfromw{
		\in@hundreds{\@p@swidth}{\@bbh}{\@bbw}
		\edef\@p@sheight{\@result}
}
\def\compute@handw{
		\if@height 
			\if@width
			\else
				\compute@wfromh
			\fi
		\else 
			\if@width
				\compute@hfromw
			\else
				\edef\@p@sheight{\@bbh}
				\edef\@p@swidth{\@bbw}
			\fi
		\fi
}
\def\compute@resv{
		\if@rheight \else \edef\@p@srheight{\@p@sheight} \fi
		\if@rwidth \else \edef\@p@srwidth{\@p@swidth} \fi
}
\def\compute@sizes{
	\compute@bb
	\compute@handw
	\compute@resv
}
\def\psfig#1{\vbox {
	\ps@init@parms
	\parse@ps@parms{#1}
	\compute@sizes
	\if@scale
                \if@verbose
                        \typeout{psfig: scaling by \@p@scale}
                \fi
                \@ScaleInHundreds{\@p@swidth}
                \@ScaleInHundreds{\@p@sheight}
                \@ScaleInHundreds{\@p@srwidth}
                \@ScaleInHundreds{\@p@srheight}
        \fi
	\ifnum\@p@scost<\@psdraft{
		\if@verbose{
			\typeout{psfig: including \@p@sfile \space }
		}\fi
		\special{ps::[begin] 	\@p@swidth \space \@p@sheight \space
				\@p@sbbllx \space \@p@sbblly \space
				\@p@sbburx \space \@p@sbbury \space
				startTexFig \space }
		\if@angle
			\special {ps:: \@p@sangle \space rotate \space} 
		\fi
		\if@clip{
			\if@verbose{
				\typeout{(clip)}
			}\fi
			\special{ps:: doclip \space }
		}\fi
		\if@prologfile
		    \special{ps: plotfile \@prologfileval \space } \fi
		\special{ps: plotfile \@p@sfile \space }
		\if@postlogfile
		    \special{ps: plotfile \@postlogfileval \space } \fi
		\special{ps::[end] endTexFig \space }
		\vbox to \@p@srheight true sp{
			\hbox to \@p@srwidth true sp{
				\hss
			}
		\vss
		}
	}\else{
		\if@draftbox{		
			\hbox{\fbox{\vbox to \@p@srheight true sp{
			\vss
			\hbox to \@p@srwidth true sp{ \hss \@p@sfile \hss }
			\vss
			}}}
		}\else{
			\vbox to \@p@srheight true sp{
			\vss
			\hbox to \@p@srwidth true sp{\hss}
			\vss
			}
		}\fi

	}\fi
}}
\def\psglobal{\typeout{psfig: PSGLOBAL is OBSOLETE; use psprint -m instead}}
\psfigRestoreAt

\newif\ifpdf
\ifx\pdfoutput\undefined
  \pdffalse
\else
  \pdfoutput=1
  \pdftrue
\fi

\ifpdf
  \usepackage[pdftex]{graphicx}
  \usepackage[pdftex]{color}
  \DeclareGraphicsExtensions{.pdf,.png,.jpg}
\else
  \usepackage[dvips]{graphicx}
  \usepackage[dvips]{color}
  \DeclareGraphicsExtensions{.eps,.epsi,.ps}
\fi

\usepackage{times}

\def\midv{\mathop{\,|\,}}

\long\def\cbk#1{{\color{red}[CBK: #1]}}
\newlength\colwidth \setlength\colwidth{3.25in}

\title{Explaining Scenarios for Information Personalization}
\author{Naren Ramakrishnan, Mary Beth Rosson, and John M. Carroll\\
Department of Computer Science\\
Virginia Tech, VA 24061\\
Email: \{naren,rosson,carroll\}@cs.vt.edu}
\begin{document}

\maketitle
\begin{abstract}
\noindent
Personalization customizes information access. The PIPE (`Personalization is
Partial Evaluation') modeling methodology represents interaction with
an information space as a program. The program is then specialized to a
user's known interests or information seeking activity by the technique
of partial evaluation. In this paper, we elaborate PIPE by considering
requirements analysis in the personalization lifecycle. We investigate the use 
of scenarios as a means of identifying and analyzing personalization 
requirements. As our first result, we show how designing a PIPE representation
can be cast as a search within a space of PIPE models, organized along 
a partial order. This allows us to view the design of a personalization 
system, itself, as specialized interpretation of an 
information space. We then exploit the underlying equivalence 
of explanation-based generalization (EBG) and partial evaluation to realize
high-level goals and needs identified in scenarios;
in particular, we specialize (personalize) an information space based 
on the explanation of a user scenario in that information space, just as 
EBG specializes a theory based on the explanation of an example in that theory.
In this approach, personalization becomes the transformation of information
spaces to support the explanation of usage scenarios. An
example application is described.\\

\noindent
{\bf Word counts:} Abstract (183 words), Main Text (10200), 
Bibliography (1092), Appendix (2375).\\

\end{abstract}

\newpage
\tableofcontents
\newpage
\section{Introduction}
Personalization constitutes the mechanisms and technologies required to
customize information access to the end-user.  It can be defined as
the automatic adjustment of information content, structure, and presentation
tailored to an individual user. With the rapid increase in the amount of
information
being placed online, the scope of personalization today
extends to many different forms of information content and
delivery~\cite{cacm-broader,cacm-kantor,cacm-streams}, not just web pages.
It is
estimated that by the year 2003, personalization services
will constitute the major component of the Internet industry~\cite{appian}.

There are undoubtedly `personal views of personalization'~\cite{cacm-personal}.
This is evident both from the numerous ways in which the term is informally
interpreted as well as the various choices available for designing, building,
and targeting personalization systems~\cite{adaptive-sites,specissue,rus}.
A simple form of personalization is where a web portal such as {\tt
myCNN.\hskip0ex com}
allows a user to customize newsfeeds, colors, and layouts to create a
personal gateway
to the Internet~\cite{manber}. This example abstracts the personalization
problem to a point where
the burden of completing the personalization task is shifted to the user,
who must specify
the settings. Another form of personalization involves a web browser that
automatically
`hides' hyperlinks that will not lead to interesting pages. This example relies
on more
sophisticated user modeling; for example browsing history may be used to
predict pages of interest.
A third example is
the recommendation facility at {\tt amazon.com} that suggests books
according to
similarities in purchase behavior.

Not withstanding this variety, a core body of personalization
algorithms and techniques have emerged.
For instance, the mining
of web user logs to identify browsing patterns has matured into a
well-abstracted data mining
problem~\cite{cacm-mulvenna}. Similarly, algorithms for determining
similarities between
buying patterns have
been studied and scaled to realistic dimensions.
However, the process of analyzing and specifying
requirements for personalization and
designing a system that achieves
the desired functional goals is still
an ill-understood and under-emphasized research issue. In fact, the {\it
lifecycle}
underlying design and deployment of personalization systems has not been
articulated well enough to enable the investigation of these issues.

It is difficult to capture specifications of requirements
independent of particular personalization algorithms or techniques. Beyond the
familiar cognitive gap between specifying and implementing requirements,
this is due to the dynamic nature of Internet technologies,
where a new development (e.g., cookies~\cite{cookies}) enables a form of
personalization that was
not possible before. Consequently, the lifecycle for personalization systems
tends to reflect the
solution-first strategy of inventing a specific technique and then implementing
it in a demonstration system. The hazards of such an approach are well
documented~\cite{jack-making-use}.

Our goal in this paper is to begin building a bridge from the high-level
design goals and
functional requirements for personalization on the one hand, and the
specific techniques and algorithms used
to realize these goals and requirements on the other. Our approach is
motivated by the recent development of a modeling methodology for
personalization
systems --- PIPE (`Personalization is Partial
Evaluation')~\cite{naren-ic,pipe-tois}.
Personalization systems are designed and implemented in PIPE by modeling an
information seeking
interaction in a programmatic representation.  PIPE helps realize
a variety of individual personalization algorithms and
enables the view of personalization as specializing representations. However,
PIPE currently supports only the interaction modeling required of a
personalization
system; it does not address earlier stages in the lifecycle of
personalization system
design (such as requirements analysis) or later stages (such as
verification and validation).

We elaborate how to integrate the use of PIPE with the early stages of the
personalization lifecycle, in particular
capturing requirements and translating them into the design of
software. In extending PIPE, we employ
scenario-based methods for analyzing and representing
usage tasks.
Scenarios and scenario-based methods are ideally suited for our purposes
because they
help identify personalization opportunities and organize design rationale
for a system in terms of its constituent facilities.
Two key contributions emerge from our approach. First, we relate PIPE to
the context
in which personalization scenarios are envisioned, abstracted, and realized
in an information
system, and thus contribute to a better understanding of the lifecycle
underlying personalization
system design. In particular, we relate personalization to the ability to
operationalize the {\it explanation} of a scenario 
of (intended) usage. Second, we provide new
techniques for
managing and reasoning with scenarios that not only aid in personalization
system design
but also find applications in other situations that involve transformation
of representations.
For instance, the construction of simplified views of systems for training
and demonstration purposes
can be expressed using these methods.

\subsection*{Reader's Guide}
The balance of the paper is organized as follows. In
Section~\ref{pipe-modeling}, we introduce
the PIPE modeling methodology for personalization. We describe its
capabilities, shortcomings, and
relate PIPE to other projects that represent and reason about
information seeking.
Section~\ref{newsec}
takes the first steps toward reasoning from scenarios (as representations
of requirements) to
modeling opportunities in PIPE. Section~\ref{explain} further
describes how scenario-based methods can extend PIPE to apply to the
earlier stages in the
lifecycle of personalization system design, such as requirements analysis
and high-level specification
of goals. Finally, 
Section~\ref{discuss}
identifies opportunities for
future research in both scenario-based methods and personalization systems
design. A case study that illustrates many of
the ideas introduced in this paper is provided in the Appendix. 

\section{PIPE: Personalization by Partial Evaluation}
\label{pipe-modeling}

As a methodology,
PIPE~\cite{naren-ic} makes no commitments to a particular
algorithm, format for information resources,
type of information seeking activities or, more basically, the nature
of personalization delivered. Instead, it emphasizes the modeling of an
information space in a way where descriptions of
information seeking activities can be
represented as partial information. Such
partial information is then exploited (in the model) by
{\it partial evaluation}, a technique popular in the programming languages
community~\cite{jones}.

\subsection{Example: Personalizing a Browsing Hierarchy}
\label{eggs}
\begin{figure}
\centering
\begin{tabular}{cc}
& \mbox{\psfig{figure=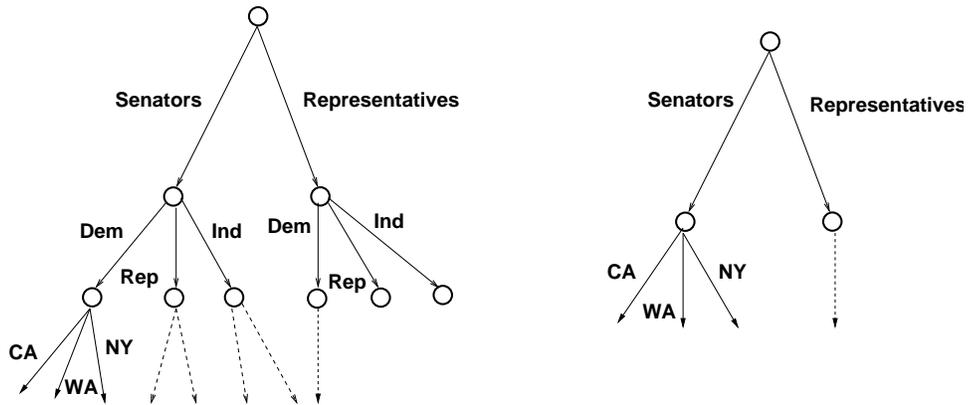,width=5in}}
\end{tabular}
\caption{Personalizing a browsing hierarchy. (left)
Original information resource, depicting information about members
of the US Congress. The labels on edges represent choices and selections
made by a navigator. (right) Personalized hierarchy with respect to the
criterion `Democrats.' Notice that not only the pages, but also their structure is 
customized for (further browsing by) the user.}
\label{senator}
\end{figure}

It is easy to illustrate the basic concepts of PIPE by describing its
application to personalizing a browsing hierarchy. Consider a congressional
web site, organized in a hierarchical fashion, that provides information
about US Senators, Representatives, their party and
state affiliations (Fig.~\ref{senator} (left)).
Assume further that we
wish to personalize the site so that a reduced or restructured hierarchy is
made available for
each user.
The first step to
modeling in PIPE involves thinking of information as being organized along
a motif
of interaction sequences. We can identify two such organizations ---
the site's layout and design that influences how a user interacts with it,
and the user's
mental model that indicates
how best her information seeking goals are specified and realized. In
Fig.~\ref{senator} (left),
the designer has made a somewhat arbitrary partition, with type of politician
as the root level dichotomy, the party as the second level, and state at
the third.
However the user might think of politicians by party first, a viewpoint
that is not
supported by the current site design. Site designs that are hardwired to
disable some
interaction sequences can be called `unpersonalized' with respect to
the user's mental model.

One typical personalization solution involves anticipating every type of
interaction sequence beforehand,
and implementing customized interfaces (algorithms) for all of
them~\cite{hearst-setting}.
For independent levels
of classification (such as in Fig.~\ref{senator} (left)), this usually implies
creating
and storing separate trees of information hierarchies. Sometimes, the site
designer
chooses an intermediate solution that places a prior constraint
on the types and forms of
interaction sequences supported. This is frequently implemented by
directing the user to one of several
predefined categories (e.g., `to search by State, click here.'). It is clear
that such solutions can involve
an exponential space of possibilities and lead to correspondingly
cumbersome site designs.

\begin{figure}
\centering
\begin{tabular}{|l|l|} \hline
{\tt int pow(int base, int exponent) \{} & {\tt int pow2(int base) \{} \\
\,\,\,\,\,{\tt int prod = 1;} & \,\,\,\,\,{\tt return (base * base);} \\
\,\,\,\,\,{\tt for (int i=0;i<exponent;i++)} &  \} \\
\,\,\,\,\,\,\,\,\,\,{\tt prod = prod * base;} & \\
\,\,\,\,\,{\tt return (prod);} & \\
\} & \\
\hline
\end{tabular}
\caption{Illustration of the partial evaluation technique.
A general purpose {\tt pow}er function written in C (left) and
its specialized version (with {\tt exponent} statically set to 2) to handle
squares
(right). Such specializations are performed automatically by partial
evaluators such as C-Mix.}
\label{pe}
\end{figure}

\begin{figure}
\centering
\begin{tabular}{|l|l|} \hline
{\tt if (Sen)} & \\
\,\,\,\,{\tt if (Dem)} & \\
\,\,\,\,\,\,\,\,{\tt if (CA)} & \\
\,\,\,\,\,\,\,\,\,\,\,\,{$\cdots \cdots \cdots$} & {\tt if (Sen)}\\
\,\,\,\,\,\,\,\,{\tt else if (NY)} & \,\,\,\,{\tt if (CA)} \\
\,\,\,\,\,\,\,\,\,\,\,\,{$\cdots \cdots \cdots$} & \,\,\,\,\,\,\,\,{$\cdots \cdots \cdots$}\\
\,\,\,\,{\tt else if (Rep)} & \,\,\,\,{\tt else if (NY)}\\
\,\,\,\,\,\,\,\,{$\cdots \cdots \cdots$} & \,\,\,\,\,\,\,\,{$\cdots \cdots \cdots$} \\
{\tt else if (Repr)} & {\tt else if(Repr)} \\
\,\,\,\,{\tt if (Dem)} &  \,\,\,\,{$\cdots \cdots \cdots$} \\
\,\,\,\,\,\,\,\,{$\cdots \cdots \cdots$} & \\
\,\,\,\,{\tt else if (Rep)} & \\
\,\,\,\,\,\,\,\,{$\cdots \cdots \cdots$} & \\
\hline
\end{tabular}
\caption{Using partial evaluation for personalization. (left) Programmatic input
to partial evaluator, reflecting the organization of information in Fig.~\ref{senator} (left).
(right) Specialized program from the partial evaluator, used to create the personalized
information space shown in Fig.~\ref{senator} (right).}
\label{sen1}
\end{figure}

The approach in PIPE is to create a programmatic representation of the
space of possible interaction sequences, and then to use the technique of
partial evaluation to realize individual interaction sequences. PIPE models
the information space as a program,
partially evaluates the program with respect to (any) user input, and
recreates a personalized
information space from the specialized program.

The input to a partial evaluator
is a program and (some) static information about its arguments. Its
output is a specialized version of this program (typically in the same
language),
that uses the static information to `pre-compile' as many operations
as possible. A simple example is how the C function {\tt pow}
can be specialized to create a new function, say
{\tt pow2}, that computes the square of an integer.
 Consider for example,
the definition of a {\tt pow}er function shown in the left part of
Fig.~\ref{pe}
(grossly simplified for presentation purposes).
If we knew that a particular user will utilize it
only for computing squares of
integers, we could specialize it (for that user) to produce the {\tt pow2}
function.
Thus, {\tt pow2} is obtained automatically (not by a human programmer)
from {\tt pow} by precomputing all expressions that involve {\tt exponent},
unfolding the for-loop, and by various other compiler transformations such as
{\it copy propagation} and {\it forward substitution}.
Automatic program specializers are available for C, FORTRAN, PROLOG, LISP,
and several other important
languages. The interested reader is referred to~\cite{jones} for a good
introduction.
While the traditional motivation for using partial evaluation is to achieve
speedup
and/or remove interpretation overhead~\cite{jones}, it can also be viewed
as a technique
for simplifying program presentation, by removing inapplicable, unnecessary,
and `uninteresting' information (based on user criteria) from a program.

Thus we can abstract the situation in Fig.~\ref{senator} (left) by the program of
Fig.~\ref{sen1} (left) whose structure models the information resource (in
this case, a hierarchy of web pages) and whose control-flow models
the information seeking activity within it (in
this case, browsing through the hierarchy by making individual selections).
The link
labels are represented as program variables and semantic dependencies
between links
are captured by the mutually-exclusive {\tt if..else} dichotomies. To
personalize this site,
for say, `Democrats,' this program is partially evaluated with
respect to
the variable {\tt Dem} (setting it to one and all
conflicting variables
such as {\tt Rep} to zero). This produces
the simplified program in the right part of Fig.~\ref{sen1}
which can be used to recreate web pages with personalized web content (shown in
Fig.~\ref{senator} (right)).
For hierarchies such as in Fig.~\ref{senator}, the representation afforded
by PIPE (notice the nesting of conditionals in Fig.~\ref{sen1}, left)
is typically much smaller than expressing the same as
a union of all possible interaction sequences.

Since the partial evaluation of a program results in another program, the PIPE
personalization operator is {\it closed.} In terms of interaction, this
means that
any modes of information seeking (such as browsing, in Fig.~\ref{sen1})
originally modeled in the program are preserved. In the above example,
personalizing a browsable
hierarchy returns another browsable hierarchy.  The closure property also
means that the
original information seeking activity (browsing) and personalization can be
interleaved in
any order. Executing the program in the form and order in which it was
modeled amounts
to the system-initiated mode of browsing. `Jumping ahead' to nested
program segments by
partially evaluating the program amounts to the user-directed mode of
personalization.
In Fig.~\ref{sen1} (right), the simplified program can be rendered and browsed in
the traditional sense,
or partially evaluated further with additional user inputs.  PIPE's use of
partial
evaluation is thus central to realizing a {\it mixed-initiative} mode of
information seeking~\cite{pipe-pepm},
without explicitly hardwiring all possible interaction sequences.

\subsection{Modeling in PIPE}
\label{factors}
Modeling an information space as a program that encapsulates the underlying
information
seeking activity is key to the successful application of PIPE. For browsing
hierarchies, a programmatic
model can be trivially built by a depth-first crawl of the site. 
In addition, a variety of other information spaces and corresponding
information seeking activities can be modeled in PIPE.
In~\cite{naren-ic,pipe-tois},
we have described modeling options for representing
information integration,
abstracting within a web page, interacting with recommender systems, 
modeling clickable maps, representing
computed information,
and capturing syntactic and semantic constraints
pertaining
to browsing hierarchies. Opportunities to curtail the cost of partial
evaluation for
large sites are also described in~\cite{pipe-tois}. We will not address
such modeling aspects here
except to say that the effectiveness of a PIPE implementation depends on the
particular modeling choices made {\it within} the programmatic
representation (akin
to~\cite{rabbit}).
We cannot overemphasize this aspect ---- an
example such as Fig.~\ref{sen1} can be made `more personalized' by conducting
a more sophisticated modeling
of the underlying domain. For example, individual
politicians' web pages at the leaves of Fig.~\ref{senator} could be modeled
by a deeper nesting of conditionals involving address, education, precinct,
and other
attributes of the individual. In other words, a single page could be
further modeled
as a browsable hierarchy and `attached' (functionally invoked) at various
places
in the program of Fig.~\ref{sen1} (left).
Conversely, the example in Fig.~\ref{senator} can be made
`less personalized' by requiring categorical information along with user input.
For instance, replacing {\tt if (Dem)} in Fig.~\ref{sen1}
with {\tt if (Party=Dem)} implies that the specification of the type of
input (namely that `Democrat' refers to the `name of the party') is required
in order for the statement to be partially evaluated.
Personalization systems built with PIPE can thus be distinguished by what
they model and the forms of customization enabled by applying
partial evaluation to such a modeling.

\subsection{Reasoning about Representations}
\label{reason}
Not all information spaces (and information seeking activities) can be
effectively modeled in PIPE. For example, a depth-first crawl of a 
site based on social network navigation
(e.g., [{\tt www.\hskip0ex imdb.\hskip0ex com}]) will result in spaghetti
code. In such cases, we
need a more complete understanding of the processes by which an online
information resource is created,
expressed, validated, and used. Even for sites that are
easily personalized, PIPE requires that they be modeled so that 
all information seeking activities are expressible as partial inputs. 
For instance, consider the following three information seeking activities 
in the context of Fig.~\ref{senator} (left).

\vspace{-0.3in}
\begin{descit}{}
\begin{description}
\item [User 1:] I will specify a party name first; then I will specify the name of
a state; finally, I will browse through any remaining links at the site. 
\item [User 2:] I would like to see the list of possible states first. So, the top level 
of the site should present me links for all the possible states. 
\item [User 3:] Show me information about the Democratic Senators of California.
\end{description}
\end{descit}

The information seeking activity of User 1 can be easily realized in the representation
of Fig.~\ref{sen1} (left), since we can partially evaluate the representation
with respect to the user's choice
of party and state. We say that the representation is well-factored for this activity and
that it is {\it personable} for this activity. However, the activity of User 2 cannot
be accomodated in Fig.~\ref{sen1} (left)
since it requires restructuring operations that are not describable as
partial evaluations. Applying partial evaluation to the representation of Fig.~\ref{sen1} (left)
can simplify interactions and allow User 2 
to make a choice of state out-of-turn. But it
cannot change the default order in which the interactions are modeled, 
which is by a 
branch-of-congress-party-state hierarchy. In this case, we say that the representation is
under-factored (for User 2's activity) and, equivalently, is {\it unpersonable} for it.
The reader should note that this doesn't mean that User 2's request can never be satisfied
in a PIPE model; see~\cite{pipe-tois} for an alternate representation of the information space
that is personable for User 2's activity (and is hence, well-factored for it).

Now, consider how we will satisfy User 3's request. 
This user has specified choices for all possible program variables --- 
involving state, party, and branch of Congress. This amounts to 
a {\it complete evaluation}, rather than a partial evaluation. Complete 
evaluation in a PIPE model
implies that every possible aspect of interaction is specified 
in advance, obviating the need for any interaction! 
Since PIPE emphasizes the specialization of interaction by partial evaluation,
the representation of
Fig.~\ref{sen1} (left) offers no particular advantages for User 3's activity.
In this case, we say that the representation is over-factored and, 
again, is unpersonable (by partial evaluation) for 
User 3's activity. Thus, both under-factorization and over-factorization
lead to unpersonable representations of information spaces.
The interesting representations are in between.

%It is difficult to create a representation that is well-factored, at once, 
%for a variety of
%information-seeking interactions. The easiest way to
%create representations --- by anticipating all possible information-seeking
%goals --- is over-factored for all interactions! For instance,
%an over-factored design for all of the above three users would be a
%web site that has the following top-level prompt:
%
%\vspace{-0.3in}
%\begin{descit}{}
%\vspace{-0.1in}
%\begin{description}
%\item click [here] if you are the user who likes to specify party name first, 
%then name of a state, and then browses through remaining links.
%\item click [here] if you are the user who would like to see list of 
%possible states first.  
%\item click [here] if you are the user who knows values for all three attributes 
%of politicians.
%\end{description}
%\end{descit}
%
%\noindent
%Such solutions bucket all users into well-defined categories by
%over-specifying the personalization problem.
In practice, it is acceptable to 
have a few situations that involve 
complete evaluation, as long as they are a small fraction of the total 
number of information seeking activities (that are to be accomodated in 
a PIPE model). 
More discussion about representations and their factorizations 
is available in~\cite{pipe-tois}; for the purposes of this paper, it suffices 
to note that we have various possibilities for representing information 
spaces in PIPE and that it is important to choose a representation that is 
well-factored.
 
\subsection{Related Research}
In name and spirit, PIPE's personalization by partial evaluation is similar
to RABBIT's~\cite{rabbit}
retrieval by reformulation. Both these approaches involve the modeling of
information seeking in a setting
that emphasizes (i) reconciling the mismatch
between how an information space is organized and how a particular user
forages in it,
(ii) closure properties of the transformation operators, and (iii)
the design of information systems in ways that highlight new evaluation
criteria. Like RABBIT, PIPE
assumes that `the user knows more about the generic structure of the
[information space] than [PIPE]
does, although [PIPE] knows more about the particulars ([web
pages])~\cite{rabbit}.' For
instance, personalization by partial evaluation is only as effective as the
ease with
which program variables could be set (on or off) based on information
supplied by the user.
As such, PIPE has no semantic understanding of the representation.

PIPE also differs from RABBIT in important ways. It emphasizes the modeling
of an information
space as well as an information seeking activity in a unified programmatic
representation.
Its single transformation operator (partial evaluation) provides a basis to
reason about the design of
personalization systems. Since partial evaluation works best for highly
parameterized and structured spaces, the PIPE viewpoint relates the 
personalizability of an information resource to the factorizability 
of its representation. A well-factored information
space is thus a personable
one, since information seeking activities are expressible as partial inputs.
%This means that for an information seeking task that can be modeled
%programmatically,
%partial evaluation can be used as a theoretical way of assessing {\it any}
%personalization system
%designed for that task, not just ones designed by PIPE.

Research at the intersection of information systems and HCI has a strong
tradition, with many other
prominent examples. Both the Scatter/Gather~\cite{scatter-gather} and
Dynamic Taxonomies~\cite{tkde-navigation}
projects rely on defining a set of operations under which transformations
made on an information
space are closed. These projects concentrate on retrieval and navigation,
respectively. While
there has been considerable research in web
personalization~\cite{adomavicius01expert-driven,
ira, fab, cacm-hirsh, grouplens, cacm-jaideep, phoaks},
many of these algorithms/systems (or in some cases their results)
are usefully viewed as modeling choices to be made in a PIPE
implementation. For instance,
the graph-theoretic recommendation algorithm described in~\cite{ira} can
be modeled as a function in PIPE, so that the results of the function are
used to set
values for program variables, which can in turn be `linked' to more
detailed information about the
recommended artifacts.
There have also
been attempts at defining theories of information access, suitable for the
design of personalization
systems~\cite{pirolli-card}. Pirolli~\cite{pirolli-chapter}
explains the idea of `information foraging' and analyzes projects such
as Scatter/Gather in this context.

Empirical research involving usage modeling
and information capture is also relevant here. Drawing ideas from the ACT-R
theory of cognition, Pirolli
et al.~\cite{forage2} describe how a quantitative model of information
foraging
can be defined. Tools for capturing history of interaction
in information foraging are also well
studied~\cite{holland-hill,footprints}. Mining
web user logs
has become a popular technique for obtaining models of site
navigation~\cite{cacm-jaideep,
cacm-mulvenna,cacm-myra}. While this strand of research
has arrived at rich, quantitative
models of site usage and navigation, there is a persistent gap between what
could
be mined from site usage and how the site could be automatically
transformed to conform
to any identified needs. Typical approaches to bridging
from the results of site usage studies
to opportunities for site restructuring
are heuristic (see for instance~\cite{cacm-myra}) and are limited in the
transformations they employ~\cite{adaptive-sites}.

\section{From Scenarios to Modeling Choices in PIPE}
\label{newsec}
As mentioned earlier, PIPE's modeling methodology requires a programmatic
representation (such as Fig.~\ref{sen1}, left) for partial evaluation.
Where do such representations come from? In this section, we analyze how
personalization requirements originate in usage contexts, and 
how they can help to build the representations of information spaces needed 
in PIPE. In this sense, we are extending the PIPE methodology `upstream' 
in the personalization system design life cycle, to include requirements 
analysis and specification.

While even a very general notion of requirements gathering applies in our 
situation~\cite{jack-require,
kieras-crc,human1,sommerville-crc}, personalization offers the unique
viewpoint of interpreting a general, existing information resource in a 
specialized manner (and thus, indirectly improving it). Studies in 
traditional IR contexts (e.g., see~\cite{human3,rabbit}) have
shown that one way to achieve such specialized interpretation is to support
the iterative reformulation of information requests. Besides reconciling
the mental mismatch between user expectations and the facilities 
afforded by an information system, reformulation engages the
user in an active dialog with the system, using both system 
features and user input to complete the information seeking activity.

We propose that such active search and reformulation episodes can be 
anticipated, revealed, and modeled by usage scenarios. These scenarios can
then form the basis of a scenario-based analysis and design (SBD)
process~\cite{jmc1,rosson-jmc}. Current practice is observed and
described in scenarios, and such scenarios help analyze how designers 
and users think about complex information resources. 
By helping to reason about the tradeoffs and design rationale associated 
with system design decisions, scenarios can aid in identifying 
opportunities for personalization. How to systematically proceed from 
high-level goals and needs identified in a scenario to a programmatic 
model in PIPE is the subject of the balance of this paper.

%As we will show later, the design rationale is a key 
%research result by itself, since it serves as a basis for understanding 
%what it means to design and use partial evaluation 
%techniques for information personalization.
%
%purposes, the design rationale can serve as a key research 
%These scenarios can then
%form the basis of a scenario-based analysis and
%design (SBD) process~\cite{jmc1,rosson-jmc}.
%SBD methods involve a combination of analytic and empirical activities ---
%current practice is observed and described in scenarios, and these
%scenarios are analyzed, transformed, refined, and
%evaluated in a continuing cycle of design and analysis. Central design products
%include narratives of realistic use that motivate the design of a system and
%an accompanying analysis of the tradeoffs or design rationale associated with
%system design decisions~\cite{jmc1}. For personalization systems, these
%scenarios and design rationale explain how designers and users think
%about complex information resources. As we will show later, the design
%rationale is a key research result by itself, since it serves as a 
%basis for understanding what it means to design and use partial evaluation 
%techniques for information personalization.

\begin{figure}
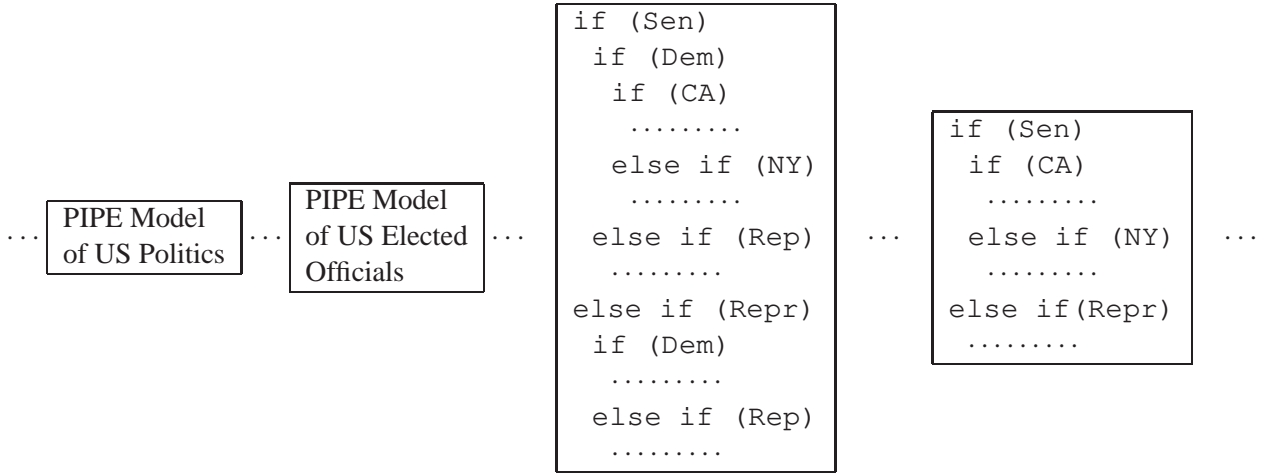

\centering
\begin{tabular}{llllll}
$\cdots$
\begin{tabular}{|l|} \hline
PIPE Model\\
of US Politics\\
\hline
\end{tabular}
$\cdots$
\begin{tabular}{|l|} \hline
PIPE Model\\
of US Elected\\
Officials \\
\hline
\end{tabular}
$\cdots$ &
\begin{tabular}{|l|} \hline
{\tt if (Sen)} \\
\,\,\,\,{\tt if (Dem)} \\
\,\,\,\,\,\,\,\,{\tt if (CA)} \\
\,\,\,\,\,\,\,\,\,\,\,\,{$\cdots \cdots \cdots$} \\
\,\,\,\,\,\,\,\,{\tt else if (NY)} \\
\,\,\,\,\,\,\,\,\,\,\,\,{$\cdots \cdots \cdots$} \\
\,\,\,\,{\tt else if (Rep)} \\
\,\,\,\,\,\,\,\,{$\cdots \cdots \cdots$} \\
{\tt else if (Repr)} \\
\,\,\,\,{\tt if (Dem)} \\
\,\,\,\,\,\,\,\,{$\cdots \cdots \cdots$} \\
\,\,\,\,{\tt else if (Rep)} \\
\,\,\,\,\,\,\,\,{$\cdots \cdots \cdots$} \\
\hline
\end{tabular}
& $\cdots$ & 
\begin{tabular}{|l|} \hline
{\tt if (Sen)}\\
\,\,\,\,{\tt if (CA)} \\
\,\,\,\,\,\,\,\,{$\cdots \cdots \cdots$}\\
\,\,\,\,{\tt else if (NY)}\\
\,\,\,\,\,\,\,\,{$\cdots \cdots \cdots$} \\
{\tt else if(Repr)} \\
\,\,\,\,{$\cdots \cdots \cdots$} \\
\hline
\end{tabular}
& $\cdots$
\end{tabular}
\caption{A space of PIPE models organized by the partial evaluation
relation.}
\label{continuum}
\end{figure}

Before we describe our approach, it is important to make some preliminary
remarks. Let us revisit the two PIPE models in Fig.~\ref{sen1}. The
model in Fig.~\ref{sen1} (right) is the result of partially evaluating
Fig.~\ref{sen1} (left) with respect to `Democrats.' However, the model of
Fig.~\ref{sen1} (left) can itself be viewed as the result of a partial
evaluation (say, of a model that provides information about
all US politicians, with respect to `congressional officials'). In other words,
Fig.~\ref{sen1} (left) is personalized for information about
members of the US Congress. Likewise, the model of Fig.~\ref{sen1} (right) 
can be viewed as the starting point of interaction with an (unpersonalized)
information system, one that is designed for people who are interested in only 
Democrats. It should thus be clear that there is actually a continuum of 
PIPE models (see Fig.~\ref{continuum}), organized along a partial order 
(where the specialization relation is partial evaluation). 

Given this observation, we can cast our requirements analysis problem as
a search within a space of PIPE models, such as Fig.~\ref{continuum}. But
we can go further. Every model in this space can be
thought of as the result of a partial evaluation or, equally, as a starting
point for partial evaluation. This means that the task of selecting
a PIPE representation for subsequent personalization 
can actually be viewed as a problem of partially
evaluating (personalizing) a more general representation! Designing a 
personalization system is thus reduced 
to a problem of personalization (of a general, and perhaps ineffective,
information space). A PIPE representation can be seen as
`freezing' some aspects
of interaction and making available some other aspects to
model users' information seeking activities. In Fig.~\ref{sen1}
(left), the model is the result of partial evaluation with respect to congressional 
officials, but program variables
pertaining to party, type, and state are available 
to represent users' personalization objectives. This viewpoint 
reinforces our idea that both designing and using personalization systems 
involve specialized interpretation of information spaces. 

Clearly this cyclic argument has to end somewhere, so what is the `starting
model' for partial evaluation? And where does it come from?
Our approach is to relate opportunities 
identified in usage scenarios to a characterization of the space of PIPE 
models, by qualifying a most-specific and the most-general elements of 
the space. We then
define an evaluation function 
to express our preference for one 
model over another. Before we can formally describe our methodology,
we must broaden our view of partial evaluation, 
moving from its algorithmic 
details as a specialization function, to larger contexts that recognize the
space of models induced by the partial order.

%our method demands that requirements identified in
%this fashion are representable as partial inputs in a PIPE model.
%This is difficult because partial evaluation is simply a program
%manipulation technique, so it does not provide any guidance on where 
%partial inputs (or the program) can come from, {\it only that they 
%are assumed given.} One way to view the challenge is as
%a need to make partial evaluation well-specified (for personalization 
%purposes), given the high-level goals and needs identified in a scenario. 
%To address this problem we must broaden our view of partial evaluation,
%moving from its algorithmic details to examples of how similar
%techniques have been applied in other contexts.

One such context is the work on explanation-based generalization 
(EBG) in AI~\cite{dejong}. Just as partial evaluation addresses the
specialization of programs, EBG addresses the specialization of domain
theories. In fact, van Haremelen and Bundy have observed~\cite{EBG_PE} that
when programs and domain theories are both represented in Prolog notation,
partial evaluation and EBG are essentially equivalent. With respect to
our scenario modeling problem, EBG makes the important addition of recognizing
the space of models induced by the partial evaluation
relation: the space is first defined by a systematic process of 
`explaining' observations and reasoning about features that are 
relevant to the observation. The vocabulary for conducting the explanation is 
provided by the domain theory; the relevant features thus identified 
help characterize the search space. Next, EBG provides a search criterion for 
evaluating models, one that emphasizes the utility and usefulness of 
the ensuing representations.

We can borrow this idea of explanation, using it to bridge from usage 
scenarios to the models and representational choices required by PIPE.
Just as an existing domain theory supports the construction of an explanation
for an example observation (which then guides the specialization of 
the theory), an existing information space can support the construction 
of an explanation for a scenario (of intended usage), which can then
guide the personalization of the information space (in our case, thus 
helping to design a personalization system).

\subsection{Explanation-Based Generalization}
To understand how we can bridge the high-level requirements uncovered
through scenarios and the programmatic modeling required by PIPE, we briefly 
review the basic ideas of EBG in an everyday context.
Consider a non-native speaker of English (Linus)
visiting the
United States. He is attempting to learn conversational constructs for
`being polite.'
The essence of EBG is that it is easier for Linus (at first) to verify or
explain why a particular conversation is
an example of politeness, than to describe or define politeness in a vacuum.
Thus, Linus observes instances of politeness and generalizes from them by
explaining
why they appear to be polite. For instance,
he witnesses the following dialog between two people:

\begin{descit}
{\bf Person 1:} Sir, I was wondering if you could point me in the direction
of Central Park.\\
{\bf Person 2:} Sure. Make a right two blocks after the gas station.
\end{descit}

At this point, Linus can infer that this is a valid example of politeness
(from Person 2's helpful response) and proceeds to explain the observation. 
By analyzing the structure of Person 1's query
and using his knowledge of how English sentences are constructed, Linus
constructs an explanation of this observation. DeJong~\cite{dejong} shows 
that an explanation can be viewed as a tree where each leaf is a 
property of the example being explained, each internal node models 
an inference procedure applied to its children, and the root is the
final conclusion supported by the explanation (namely, that the above was
an example of politeness). The explanation tree proves that
the conversation is polite and 
helps separate out the 
relevant and incidental parts of the above conversation; any attribute of 
the conversation that does not participate in the `proof' does not contribute
to politeness.

Using this structure (and his knowledge of English), Linus can then study 
how it can be generalized to other situations.
For instance, he can reason that the
phrase `Sir, I was wondering if' is what confers politeness onto the whole
sentence. He can also conclude that `Central Park' is not a 
property of politeness per se, but a feature of the
particular request. Notice that Linus could have arrived at the phrase
`Sir, I was wondering if' himself (without the above example), but that
would have required
a lot of imagination (computation, for an AI system~\cite{russel-norvig})
on his part. This is the
essence of EBG --- namely that we don't `actually learn anything factually
new from the instance~\cite{russel-norvig}' but such examples
point us in the direction in which to specialize our unmanageable domain
theory (in this
case, Linus's rules of grammar and knowledge of how English sentences are
constructed).

The reader might notice a disconnect between the G in EBG (which stands for
{\it generalization}) and our statement that EBG is really a technique for
{\it specializing}
domain theories. This can be understood by noticing that the most common
usage of EBG is in
learning concept descriptors from individual examples~\cite{dejong}. Thus,
while it is the domain theory that is being specialized, the example is 
being generalized by throwing away 
parts of its explanation structure. In other words, the domain theory
constitutes the prior knowledge
that is useful for generalization~\cite{russel-norvig}. Linus can then
exhibit his newly acquired politeness in
a different situation such as: `Sir, I was wondering if you could hold open
the elevator for me.' Or
even further, he might generalize `Sir' to include `Madam' and `Lady.'

In concept learning, 
the level to which Linus generalizes a particular explanation is influenced 
by {\it operationality}. For instance, 
if he doesn't generalize beyond `Sir, I was wondering if you could point me 
in the direction of,' then his learning can only be applied to, say, 
situations when he is lost. At the other extreme, Linus might reason that there are many other ways of
being polite (such as
`Could you please ...?') and conclude that
any well mannered phrase prefixed to a request constitutes an instance of
politeness. Such
an over-generalization is however less operational, since it assumes that
Linus has some other way of
deciding what makes a phrase `well mannered.' Operationality is thus related
to the utility of the induced generalization.

%\begin{table}
%\centering
%\begin{tabular}{|lcl|} \hline
%Learning to be polite & = & Concept Learning (in EBG) \\
%                        & = & Program Specialization (in PE)\\
%			& & \\
%Rules of Conversational English & = &  Domain Theory (in EBG) \\
%			& = & Original Program (in PE) \\
%			& & \\
%%Central Park Conversation & = & Example of Goal Concept (in EBG) \\
%%			& = &
%%\fbox{\,\,\,\,\,\,\,\,\,\,\,\,\,\,\,\,\,\,\,\,\,\,\,\,\,\,\,\,\,\,\,\,\,\,\,\,}
%%(in PE) \\
%%			& & \\
%Essential Aspects of Central Park  Conversation & = & Features of Training
%Example (in EBG) \\
%& = & Values for Static Variables (in PE) \\
%			& & \\
%Rules for Politeness    & = & Reformulated Goal Concept (in EBG) \\
%                        & = & Specialized Program (in PE) \\ \hline
%\end{tabular}
%\caption{A comparison of explanation-based generalization (EBG) and partial
%evaluation (PE).}
%\label{ebgispe}
%\end{table}

%With these observations, we can make a mapping between EBG and partial
%evaluation
%(see Table~\ref{ebgispe})
%by cross-referencing terms and concepts between the two areas. We mention
%that our mappings are
%more conservative than the stronger claims made in~\cite{EBG_PE} but the
%latter work is presented in
%the context of a first-order logic representation for the domain theory. 

\begin{figure}
\centering
\begin{tabular}{|ll|} \hline
{\bf Most general:} & $\prec$polite conversational construct$\succ$.\\
& ....\\
&$\prec$well-mannered phrase$\succ$ $\prec$Linus's desire$\succ$.\\
& ....\\
& $\prec$address$\succ$, I was wondering if $\prec$Linus's desire$\succ$.\\
& ....\\
& Sir, I was wondering if $\prec$Linus's desire$\succ$.\\
& ....\\
{\bf Most specific:} & Sir, I was wondering if you could point me in the direction of Central Park.\\
\hline
\end{tabular}
\caption{Example generalizations for the Central Park conversation.}
\label{linusgen}
\end{figure}

\subsection{Using EBG in Personalization}
\label{ebginpe}
Keller~\cite{keller-op} shows how we can think of EBG as a search 
through a concept description space such as Fig.~\ref{linusgen}.
The operationality consideration is then the objective function
used to evaluate entries in the concept description space. 
The most specific construct simply records the conversation 
and can only be replayed in an exactly similar situation. The effort to 
instantiate the construct is thus minimal but many such constructs will 
likely be needed to cover a realistic set of situations.
The most general construct involves no learning on Linus's part and 
merely restates his desire to learn polite constructs. If Linus adopts this
construct, he can have one single explanation structure to support
all situations but he has to expend the effort to instantiate it (effectively,
constructing the proof) every time he needs to be polite.

Analogously, the goal of obtaining a PIPE model is viewed as
search through a space of possible PIPE models, ordered by the partial 
evaluation operator. 
Explaining a user's successful interaction at a site (or collection of sites)
with respect to a domain theory (more on this later) will help identify 
the parts of the interaction that contribute to achieving 
the personalization objectives. The explanation tree thus serves to define 
the search space of PIPE models. The operationality boundary is then the 
point at which certain parameters are fixed in our representation of the 
information space and certain others are available to model users' 
information seeking interactions. 
%For instance, the operationality boundary in the example given in
%Section~\ref{eggs} is our requirement that the personalization system 
%cater to information about members of the US Congress {\it only.} Stated 
%alternatively, operationalization here is intended to support the subsequent
%partial evaluation of the information space with respect to the type, 
%party, and state of political officials; the representation is
%hence parameterized in terms of these variables.

\begin{figure}
\centering
\begin{tabular}{|ll|} \hline
{\bf Inputs}&---- Functional Description of Personalization Problem \\
& \,\,\,\,\,\,\,\,\,(i.e., a non-operative definition)\\ 
 & ---- Domain Theory \\
& \,\,\,\,\,\,\,\,\,(describing site layout, task models, browsing semantics; \\
& \,\,\,\,\,\,\,\,\,must support the construction of an explanation) \\
& ---- Usage Scenario  \\
& \,\,\,\,\,\,\,\,\,(showing successful realization of personalization goal;\\
& \,\,\,\,\,\,\,\,\,typically a sequence of interactions which achieves goal)\\
& ---- Operationality Criterion\\
& \,\,\,\,\,\,\,\,\,(helps evaluate alternative PIPE models;\\
& \,\,\,\,\,\,\,\,\,outlines which aspects of explanation structure can be fixed,\\
& \,\,\,\,\,\,\,\,\,and which should be available for capturing interactions with
users)\\
{\bf Output}&---- PIPE model (that defines a personalization system) \\
\hline
\end{tabular}
\caption{Formalizing requirements analysis in the personalization lifecycle
as an EBG problem.}
\label{funcos}
\end{figure}

\begin{figure}
\centering
\begin{tabular}{cc}
\mbox{\psfig{figure=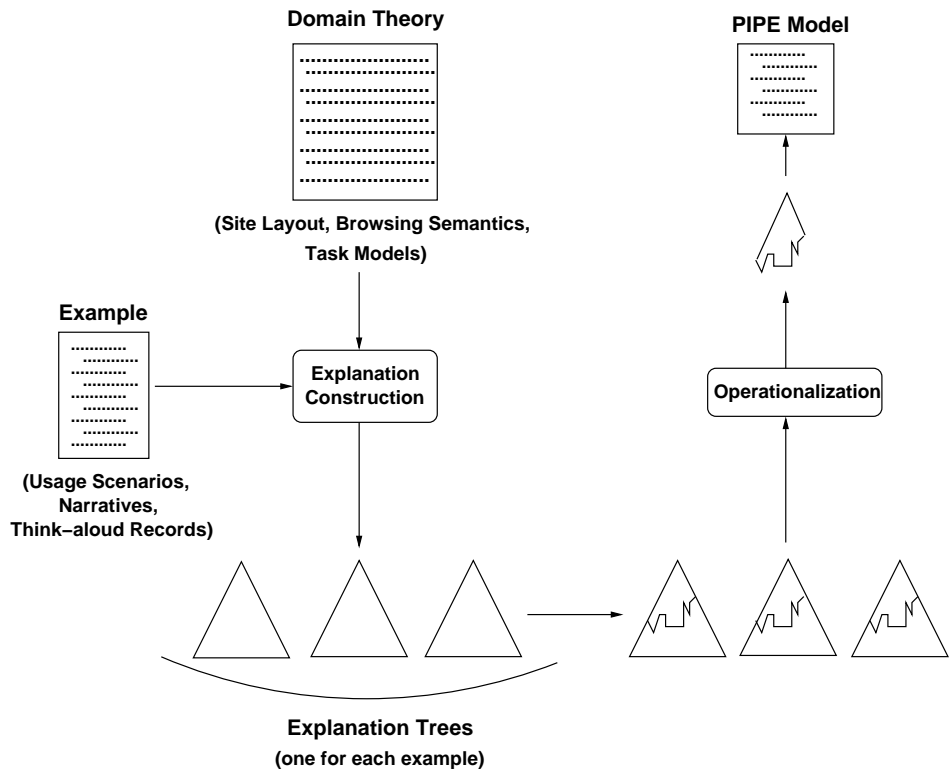,height=4in}}
\end{tabular}
\caption{From scenarios to modeling choices in PIPE: toward a lifecycle of personalization
system design.}
\label{lifecycle}
\end{figure}

Adapting from~\cite{ebg-alternate}, we can formalize requirements analysis
for PIPE as shown in Fig.~\ref{funcos}.
Our methodology requires the specification
of four inputs. A functional description of the personalization problem is
assumed so that we can distinguish between scenarios where the user was
successful in achieving his objectives from those where he was not successful.
The domain theory is the most critical aspect of the methodology and encodes
knowledge about site layout, browsing semantics, task models, and any other
information that is relevant for reasoning about the user's
personalization objectives. In addition, the domain theory language
should support inference procedures (e.g., deduction, rewriting) that 
enable the construction of explanations. Usage scenarios, narratives, and think-aloud records constitute
the third input and together with the domain theory, drive the 
explanation construction process. Finally, operationality serves as the
criterion for evaluating the space of models induced by generalizing an
explanation.

Procedurally, our methodology consists of a three-stage approach 
(see Fig.~\ref{lifecycle}):
(i) constructing explanations from scenarios of use; this reveals how 
a given site (or collection of sites) helps the user to complete his 
information seeking activity, (ii) operationalizing the explanations in 
terms of site facilities; this allows us to assess the most relevant aspects 
of the explanation structure that we would like to retain and express 
in a personalization system, and (iii) expressing the operationalized 
explanation in a PIPE model (which will allow its subsequent personalization 
for future users). We now proceed to study these steps in greater detail.

\section{Explaining Personalization as Partial Evaluation}
\label{explain} \subsection{Constructing Explanations from Scenarios} The SBD process begins with an analysis of current usage practices. This would typically take place through field work that includes observation of work sessions, interviews or surveys of domain
experts, and collection and analysis of work artifacts. The goals and
concerns of current use are synthesized and contextualized as problem
scenarios.

\begin{figure}
%\centering
%\framebox{
\begin{tabular}{|l|}\hline
\begin{minipage}{2\colwidth}
%\small
\begin{description}
\item [Nancy's scenario]: Nancy, a resident of 
North Carolina, is interested in determining the committees that 
the junior senator from her state is a member of.
She uses the PoliticalInfo web site to perform her information 
seeking activity. The top level of the site features various categories
of political information, organized according to the offices of government
(such as `President,' `Congress,' and `State Offices'). She clicks on
`Congress' and reaches a page that prompts her to choose a state. Upon
selecting `North Carolina,' the site refreshes to a selection involving
branch of Congress (Senate or the House of Representatives). Nancy selects
`Senate,' and the system now requests information on whether the senator
occupies the junior or senior seat.
By mistake, 
she clicks on an advertisement banner for campaign finance reform, which 
causes a new browser window to be opened up, soliciting information from 
Nancy for an opinion poll. She hurriedly closes
this new window, goes back to her browsing session in progress, and clicks
on `Junior Seat.' Nancy scrolls down the displayed homepage of the individual,
eyeballs the various headings, and finally spots the information about
committee memberships. She notes that there are four committees that
the senator is a member of. Satisfied with the results of her information
seeking, Nancy exits from her browser program.
\end{description}
\end{minipage}
\\\hline
\end{tabular}
\caption{Fictitious narrative of a problem scenario synthesized from field observation.}
\label{nancy}
\end{figure} 

\begin{figure}
\centering
\begin{tabular}{|rrl|} \hline
R1: & politicalinfo($x$) & $\Leftarrow$ complete($x$) \\
R2: & complete($x$) & $\Leftarrow$ officeselect($x$,``Congress'') $\wedge$ member($x$) $\wedge$ aspect($x$) \\
R3: & complete($x$) & $\Leftarrow$ officeselect($x$,``President'') $\wedge$ aspect($x$) \\
$\cdots$ & & \\
R25: & member($x$) & $\Leftarrow$ representative($x$) \\
R26: & member($x$) & $\Leftarrow$ senator($x$) \\
$\cdots$ & & \\
R32: & senator($x$) & $\Leftarrow$ branchselect($x$,``Senate'') $\wedge$ stateselect($x$,$s$) $\wedge$ seatselect($x$,``Junior Seat'') \\
R33: & senator($x$) & $\Leftarrow$ branchselect($x$,``Senate'') $\wedge$ stateselect($x$,$s$) $\wedge$ seatselect($x$,``Senior Seat'') \\
$\cdots$ & & \\
R48: & aspect($x$) & $\Leftarrow$ aspectselect($x$,``Education'') \\
R49: & aspect($x$) & $\Leftarrow$ aspectselect($x$,``Committee Memberships'') \\
R50: & aspect($x$) & $\Leftarrow$ aspectselect($x$,``Home City'') \\
$\cdots$ & & \\
S1: & stateselect($x$,$s$) & $\Leftarrow$ clickedon($x$,$p$,$s$) $\wedge$
congresslevel($p$) $\wedge$ hyperlink($s$) \\
S2: & adselect($x$,$a$) & $\Leftarrow$ clickedon($x$,$p$,$a$) $\wedge$
advertisement($a$) \\
$\cdots$ & & \\
%L1: & $\forall x,p,l,q$ follow($x$,$p$,$l$) $\wedge$ graph($p$,$l$,$q$) $\Rightarrow$ visited($x$,$q$) \\ 
%L2: & $\forall x,p,l$ follow($x$,$p$,$l$) $\Rightarrow$ clickedon($x$,$l$) \\
%$\cdots$ & \\
%N1: & $\forall x1,x2,p,l,q$ follow($x1$,$p$,$l$) $\wedge$ advertisement($l$)
%$\wedge$ graph($p$,$l$,$q$) $\wedge$ newsession($x1$,$x2$) $\Rightarrow$
%visited($x2$,$q$) \\
%$\cdots$ & \\
\hline
\end{tabular}
\caption{An example domain theory for reasoning about interactions at
the fictitious PoliticalInfo site. All variables are assumed to be
universally quantified (syntax adapted from~\cite{dejong}).}
\label{egdomain}
\end{figure}

\begin{figure}
\centering
\begin{tabular}{|ll|} \hline
F1: & officeselect($x47$,``Congress''). \\
F2: & stateselect($x47$,``North Carolina'').\\
F3: & branchselect($x47$,``Senate'').\\
F4: & adselect($x47$,``Campaign Finance Reform Adverstisement'').\\
F5: & seatselect($x47$,``Junior Seat'').\\
F6: & aspectselect($x47$,``Committee Memberships'').\\
\hline
\end{tabular}
\caption{Description of Nancy's scenario for subsequent explanation.}
\label{nancy2}
\end{figure}

\begin{figure}
\centering
\begin{tabular}{cc}
\mbox{\psfig{figure=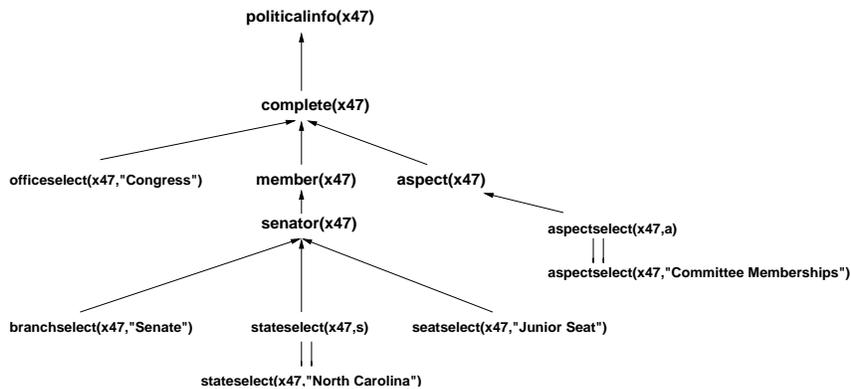,width=4.5in}}
\end{tabular}
\caption{Constructing an explanation for Nancy's scenario.
Following DeJong~\cite{dejong}, the arrows indicate
influence of rule antecedents on consequents and the parallel
lines indicate unification constraints.}
\label{nancy-tree}
\end{figure}

As an example, consider the analysis and development of scenarios for 
personalizing information about US political officials.
We might observe users' interactions with various web sites
for several hours, then interview them and analyze their decisions or 
other artifacts that they generate. Empirical techniques for
analyzing web logs and for the automatic sessionizing of
user access patterns might also be used
(see~\cite{jaideep-kais} for methods of data preparation and
gathering for this activity).
The analysis might point to a recurring scenario in which users 
select a particular site, navigate the pages in the
site by making various selections on the category of the political 
official and finally, lookup individual web pages to obtain 
specific information about a particular official. For instance,
Fig.~\ref{nancy} describes a positive instance of our modeling goal
(i.e., personalizing information about political officials at the PoliticalInfo
web site).

We now turn our attention to the domain theory which consists of: 
a modeling of the information seeking goal in terms of an underlying 
schemata; our understanding of PoliticalInfo's layout (its site structure 
and how link choices specify information seeking attributes), and;
%constraints on interaction that can be inferred from how the political 
%system is structured (e.g., `every state has only two senators but the
%number of representatives varies'), and;
aspects that capture how browsing interactions take place at the site (such
as `clicking on advertisement banners cause new browser windows to
be opened up'). 
While EBG is sometimes proposed as a possible computational model of human
concept formation and cognition, it is important to note that the purpose 
of a domain theory in our methodology is {\it not} to explain
users' behavior at this level. Rather, we seek to encode information in the 
domain theory that helps relate interactions with information systems to 
the realization of information seeking objectives. This requires that
the domain theory be sufficient to reason deductively why an observed
sequence of interactions achieves the desired personalization objectives. Our 
experience is that for a variety of information spaces that support a 
goal-oriented view of information seeking 
(e.g., browsing hierarchies involving taxonomic relationships), this 
assumption of the availability of a domain theory can indeed be satisfied. 

%, semantics of 
%how attributes of political officials relate to one another (e.g., `choice of
%state and choice of branch of US congress are independent of each other'),
%taxonomic and cardinality constraints (e.g., `all states have only two senators,
%but the number of representatives can vary from one in a state like
%South Dakota to 52 in a state like California'), and aspects that capture 
%how browsing interactions take place (such as the fact that new
%browser windows can be opened and particular spatial locations on maps can be
%clicked). 

For instance, a domain theory for interactions at PoliticalInfo can be organized
as shown in Fig.~\ref{egdomain}.
The first part of the theory describes an underlying schema for achieving
the politicalinfo information seeking goal. Rule R1, in particular, is 
our functional specification of the personalization problem. It is 
non-operative and merely states that an interaction session ($x$) that is 
complete can help satisfy the personalization objectives. Rules R2 and R3 
describe two specific ways in which an
interaction can be complete, namely by either
concentrating on aspects dealing with members of the US Congress or 
on the President. Members of the US Congress, in turn, are defined as
either senators or representatives (rules R25 and R26). Notice that there
are many possibilities for defining member($x$) and these are encoded
in the domain theory.
The second part of the domain theory describes 
how primitive actions carried out by the user correspond to 
the specification of information seeking attributes. For instance, the 
selection of a state (rule S1) can be made by clicking on an
available hyperlink from the congresslevel page. Similarly selections
of advertisements can be made by clicking on advertisements from any page
(rule S2).

Using the predicates in the domain theory, we can represent Nancy's scenario
($x47$) as shown in Fig.~\ref{nancy2}. For ease of presentation, 
Fig.~\ref{nancy2} describes only the high-level selections inferred from
Nancy's interactions instead of assertions
at the level of clicks and hyperlinks. We now proceed to show
that Nancy's scenario is a `correct'
example of accessing information about political officials. In other
words, we attempt to prove that politicalinfo($x47$) is true.
The explanation tree constructed by 
resolution is shown in 
Fig.~\ref{nancy-tree} and identifies
the salient aspects of the scenario that 
contribute to realizing Nancy's information seeking goals. In particular,
rules R26 and R32 have been used to prove that Nancy's selection of attributes 
defines a particular political official. The explanation tree also reveals
that the interactions relating to the advertisement for
campaign finance reform do not contribute to Nancy's objectives. 
We exclude the aspect of Nancy recording the details of the 
committee memberships; her satisfaction with the information seeking 
activity is assumed to be implicit in the completion of the proof.

%\begin{figure}
%\centering
%\begin{tabular}{cc}
%\mbox{\psfig{figure=finance.epsi,width=3.5in}}
%\end{tabular}
%\caption{Using a scenario editor to define and
%write narratives for Bill's personalization scenario.}
%\label{bill-scenario}
%\end{figure}
%
\subsubsection{A Note about Domain Theories}
Before we describe the next stage in our methodology, it is pertinent to make
some observations about the domain theory. First, we are not constrained to
a predicate logic representation for the domain theory. The only 
requirement is that `the representation language support the
construction of an explanation'~\cite{dejong}. Second, alternate domain
theories might permit the explanation of the same scenarios, but in
qualitatively different ways. 
This is a useful feature since it 
%out the interconnection between a usage scenario and the domain theory and
prevents overgeneralizing from the observed features of the scenario.
In addition, it allows us to compare and contrast domain theories
and determine if the resulting explanations are acceptable. Third,
we can start with a `coarse' domain theory and revise it 
by focusing on particular scenarios and situations~\cite{restructure-theories}. 
Such {\it theory revision} research is an active area of EBG, where 
explanation-based techniques are augmented with more empirically-based 
methods to address the problem of imperfect prior knowledge. Finally,
while the availability of a general schemata 
aids in the construction of a domain theory, external
guidance (from users and think-aloud records) can support the construction
of an explanation and augment domain theories that are incomplete. 

This last feature is especially useful when we extend our approach to more
complex situations, such as multiple information resources. Consider 
a scenario where Nancy is seeking information about financial investments.
During analysis and design, some sub-goals and decisions can be
inferred by think-aloud protocols while Nancy forages in 
various web sites to address her financial analysis goal. For instance, Nancy 
might report that she conducted a mental calculation of dollar amounts from
Euro currency in analyzing some merger stocks, and hence we might model a
procedure for unit conversion as an intermediate goal in our domain
theory. Similarly, Nancy might have performed a manual information
integration by copying text from one browser window to another or might
have performed a mapping from company names (e.g., `Microsoft')
to ticker symbols (`MSFT'). We would represent these as unification 
constraints or semantic mappings in our domain theory, respectively. By
augmenting a domain theory in this manner, we can summarize scenarios
collected as field data into appropriate explanation structures.

We also recognize the possibility that a domain theory might
be ineffective in explaining scenarios. Consider the case when
Nancy is seeking information about the `Democratic senator from
North Carolina' and is unsure if the senator occupies the junior or senior 
seat. If the site does not allow the direct specification of her request,
Nancy might resort to trying both choices of seats to determine 
the one occupied by the Democrat. An ineffective domain theory might 
incorrectly infer that Nancy's information seeking was 
focused on both senators! We thus need to discount some steps in the
scenario as being tentative or exploratory. This is a well studied problem 
in EBG and various strategies for reducing dependence on such `brittle
theories' (such as induction over explanations) 
have been proposed~\cite{dejong,flann}. 

The reader will also note that the explanation in Fig.~\ref{nancy-tree}
(or the domain theory) does not capture the order in which the 
attributes were specified in Nancy's scenario. In this particular
instance, the temporal sequencing of subgoals is not critical to completing
the explanation; Nancy could have selected `Senate' first (if the site
allowed it) before the choice of state was made.
In a different application, the domain theory might need to support the
construction of explanations that recognize the ordering of interactions.

\subsection{Operationalizing Explanations}
Fig.~\ref{nancy-tree}'s explanation of Nancy's scenario, while
identifying relevant parts of the domain theory, is too specific to be
used as the basis of a personalization system.
The next step in our methodology is thus to determine the parts of the
explanation structure that we would like to retain and express in a PIPE
model. A trivial step of identity elimination is first done to eliminate 
dependence on the particular scenario of Nancy (i.e.,
the $x47$ in Fig.~\ref{nancy-tree} is replaced by just $x$).
Operationalization can then be thought of as drawing a cutting plane
through the explanation tree. Every node below the plane is too specific
to be assumed to be part of all scenarios. The structure above the plane
is considered the persistent feature of all usage scenarios and is
expressed in the personalization system design. The user is then
expected to supply the details of the structure below the plane so that
the proof can be completed.

\begin{figure}
\centering
\begin{tabular}{|ll|} \hline
%\begin{descit}{} 
%\begin{description}
click [here] & if you are the user who seeks information about committee \\
& memberships of the junior senator from North Carolina. \\
click [here] & if you are the user who likes information about the \\
& educational background of the President. \\
click [here] & if you are the user who seeks details about bills proposed by the\\
& Republican representative from Littletown constituency of Virginia.\\
click [here] & $\cdots$ \\ 
\hline
%\end{description} 
%\end{descit}
\end{tabular}
\caption{Operationalizing multiple explanations at the leaf level leads to
an over-factored representation in PIPE.}
\label{over-factor}
\end{figure}

\begin{figure}
\centering
\begin{tabular}{cc}
\mbox{\psfig{figure=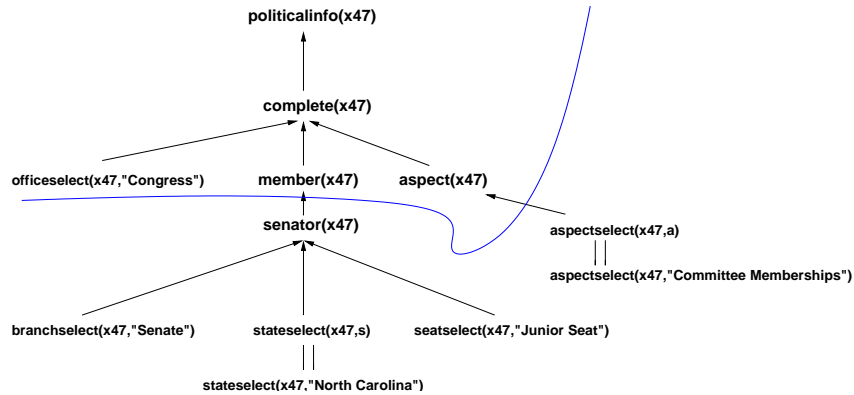,width=4.5in}}
\end{tabular}
\caption{Operationalizing the explanation for Nancy's scenario.}
\label{nancy-tree-op3}
\end{figure}

For instance, if we draw the cutting plane just below politicalinfo($x$)
then this is equivalent to no personalization 
at all. A single explanation tree can accommodate all possible 
information seeking activities but really provides no support as a 
personalization system.

If the cutting 
plane is drawn to include the leaves, then this amounts to freezing 
all aspects of Nancy's scenario so that it can be replayed in full. In such 
a case, it is unlikely that a personalization system modeled after one
explanation tree will satisfy all users. We could freeze many more such
explanation trees and the design of
the personalization system then reduces to providing a top-level prompt for
the correct explanation (see Fig.~\ref{over-factor}).
This solution anticipates all forms of interactions and 
over-specifies the personalization problem.
As is well-known in EBG, such a design
is inefficient since a new user has to search for the
correct explanation that is appropriate for his information seeking
activity. From Section~\ref{reason}, we also know that such a design would 
be over-factored and unpersonable under PIPE. This
is because the resulting PIPE model has only one argument (namely, the
choice of the correct explanation) and all invocations of such a model have
to involve complete evaluation!

Fig.~\ref{nancy-tree-op3} describes an intermediate solution where some aspects
of the explanation are fixed but some other aspects are available for 
addressing users' interactions. This operationalization induces
a system that personalizes information about
congressional officials. It
assumes that, like Nancy, a new user will invoke rule R2 from Fig.~\ref{egdomain} but, unlike Nancy, could be interested in other members of Congress
besides senators. In addition,
the part of the tree specifying the aspect of interest is also available
for specification by the user. The reader should note that such an 
operationalization will not cover scenarios where the user is interested in,
say, information about the President. To accommodate this case, we could either
move our operationality boundary or create and operationalize
another explanation tree (for
an appropriate scenario). Studying the tradeoff between these two possibilities 
constitutes the crux of operationality research. For the purposes of
this article, it suffices to note that two dimensions of operationality
are: how many explanation trees are operationalized, and 
where the boundaries are drawn in each tree.

EBG's ability to induce general constructs by explaining 
scenarios can be a drawback as well as an advantage.
If we generate a lot of templates,
then users' interactions with the personalization system can 
get burdened by a mushrooming of choices. At the same time, EBG provides
a systematic way to cluster the space of users and to determine dense regions
of repetitive interactions that could be supported. A case in point is
a web site such as [{\tt amazon.com}] that distinguishes between returning 
customers and new customers. A top-level prompt at the site makes this
distinction (sometimes, this is automated with cookies) and transfers are made
to different interaction sequences, based on the results of this choice.
For returning customers, questions about mode of payment and mailing address
are skipped because the parts of the proof dealing
with those aspects are already subsumed in the design of the system.
Another example is a web site that provides links from the top-level
page to `the top 10 frequently accessed pages at our site.' In this case, 
popular explanations have been operationalized at the leaf level and 
presented so that new users can directly access them. 
%Our methodology can thus aid in the creation of such personalized views 
%of systems for users.

%\subsubsection*{Extensions of the Basic Idea}
Operationalization is only one way to generalize an explanation to other
situations. A variety of other generalization approaches are prevalent
in the EBG literature. There are techniques that conduct generalization across 
multiple explanation trees simultaneously, by identifying recurring 
subtrees~\cite{flann}. To some extent, this can help overcome sensitivity
to initially explained scenarios and also address shortcomings in the
representation of the domain theory. There are also approaches that model and 
generalize temporal interactions 
and ones that help acquire iterative concepts. Iterative concepts are useful, 
for instance, when users to our PoliticalInfo site seek multiple aspects
of information about a political official. Nancy
was interested in only committee memberships but a different user could
have been interested in committee memberships as well as the educational
profile. Generalizing to $n$ such aspects can be achieved by acquiring
an iterative concept. We can also generalize to acquire
recursive formulations; this is useful if information seeking has
an exploratory nature to it. Linus might have visited a sports site
four times in a single scenario, whose correct generalization could be
`keep visiting the page to see if the scores have been updated.' Finally, 
since the choice of operationality is primarily driven by empirical and 
usability concerns, a variety of existing methodologies for utility 
analysis and estimation can be employed here~\cite{dejong}.

\begin{figure}
\centering
\begin{tabular}{|ll|} \hline
&$\cdots \cdots \cdots$\\
L1: &{\tt if (Senator)}\\
L2: &\,\,\,\,{\tt if (JuniorSeat)} \\
L3: &\,\,\,\,\,\,\,\,{\tt if (NC)} \\
L4: &\,\,\,\,\,\,\,\,\,\,\,\,{\tt if (CommitteeMemberships)} \\
L5: &\,\,\,\,\,\,\,\,\,\,\,\,\,\,{\tt /* Details about committees */} \\
&\,\,\,\,\,\,\,\,\,\,\,\,\,\,{$\cdots \cdots \cdots$} \\
&\,\,\,\,\,\,\,\,\,\,\,\,{\tt else if (Education)} \\
&\,\,\,\,\,\,\,\,\,\,\,\,\,\,{$\cdots \cdots \cdots$} \\
&\,\,\,\,{\tt else if (SeniorSeat)} \\
&\,\,\,\,\,\,\,\,\,\,\,\,{$\cdots \cdots \cdots$} \\
&$\cdots \cdots \cdots$\\
&{\tt else if (Representative)} \\
&$\cdots \cdots \cdots$ \\
\hline
\end{tabular}
\caption{Designing a PIPE representation from the
operationalized explanation in Fig.~\ref{nancy-tree-op3}.}
\label{pipecone}
\end{figure}

\subsection{Designing a PIPE Representation}
The last step is to express the operationalized explanation in a PIPE
representation. We can think of this stage as designing an information system
that provides all the necessary facilities to complete the proof. 
The part of the proof {\it above} the cutting plane is to be performed by 
the system, whereas the user has to supply the details of the proof 
{\it below} the cutting plane (in our case, for member($x$) and
aspect($x$)).

Ideally, the user should be able to supply her part of the proof
in as expressive a manner as possible. For instance, 
just saying `North Dakota' and `Representative' in the current political 
landscape defines a unique member of Congress. To achieve this effect 
in a PIPE model, we must ensure that all possible ways of completing the
proof are describable in terms of interaction sequences. Alternatively, we
might choose to support only certain possibilities of completing the
proof. The PIPE model should thus be parameterized in terms of
variables that help define member($x$) and aspect($x$). A representation
in C is shown in Fig.~\ref{pipecone}. Lines L1, L2, and L3 help define
member($x$) for Nancy's scenario and Line L4 helps define aspect($x$).
Since PIPE representations can be partially
evaluated, the user can specify the underlying attributes of the proof in any
order.

It is important that we also model terminal code that gets triggered upon
completion of the proof (or subproofs). In our running example, the 
terminal code (e.g., line L5) represents the results presented to Nancy 
upon successful completion of the proof (i.e., information about
committee memberships). 

The reader should recall that we have a variety of modeling options
for creating the PIPE representation. Fig.~\ref{pipecone} models the
interaction as a browsing hierarchy, similar to Fig.~\ref{sen1}. Instead we
could have modeled the interaction as a sequence of forms to be filled by
the user. 
Our representation also assumes that we have only one operationalized
explanation structure. If we have multiple explanation structures, an extra
program variable can be introduced that identifies the
explanation a given user is interested in. The PIPE model in this 
case would be a sequence of representations such 
as Fig.~\ref{pipecone}, joined together by a top-level
{\tt switch} construct. A complete description of an example application
developed by our methodology is given in the Appendix.

Programmatic PIPE models obtained by our methodology can be viewed as 
compact representations of all pertinent scenarios and, in this sense, 
are more expressive than scenario grammars~\cite{hsia} and scenario
schema~\cite{potts}. Using program compaction techniques~\cite{debray-toplas,
pipe-tois}, we can further curtail the explosion of scenario possibilities.
Any program analysis technique can then be applied to aid in scenario
analysis. For example, the technique of {\it program slicing}~\cite{slicing}
can help reason about the program parts that will be affected by changes
in a given scenario. 
By comparing these effects to those deduced before
the change, we can reason about the orthogonality of scenarios.

\subsection{Related Research}
As narrative descriptions of use, scenario-based methods have become prevalent
in various applications, including requirements
analysis~\cite{jack-require,jarke,decl-scenarios} and user interface
design~\cite{jack-making-use}.
The design of software systems from scenarios (as opposed to purely functional,
solution-first specifications) is the cornerstone of our approach. Most
relevant to our presentation is Potts's distinction between inducing scenarios
(from interaction) and deducing scenarios (from specifications)~\cite{potts}.
The operational definition of goal-achieving actions emphasized
in~\cite{potts}
is similar to our explanation structure. For instance, Potts employs a goal
hierarchy
where leaves are associated with user actions and which serve to
operationalize
goals. In addition, we are able
to mechanically transform an information space using attributes of such an
explanation
structure (by partially evaluating the programmatic representation of the
operationalized
explanation).
Other applications in automated software
engineering~\cite{hall-se} and information pattern
extraction~\cite{scenario-pattern-extract}, while supporting explanation-based
views of scenario analysis, have not been connected to the partial
evaluation aspect, which is critical for the
PIPE methodology of personalization.

The PIPE approach is also related to the use of task models in
software design. Traditionally, such integration has been achieved by
symbolic modeling techniques, motivated by object oriented (OO)
design~\cite{intTaskObj}
and languages such as UML~\cite{jack-editorial}. More recent efforts in
personalization applications can be found
in~\cite{li-catalog,schwabe2,human1,schwabe1}.
In~\cite{schwabe2}, the
derivation of models of interaction from use cases is presented.
Kramer et al.~\cite{human1} emphasize the importance of task analysis and
advocate end-user analysis of algorithms and tools employed in
personalization systems.
In~\cite{schwabe1}, the authors emphasize
an OO modeling of an information system, where personalization is introduced
as a function from the conceptual design stage.
PIPE's support for personalization, on the other hand, is built into the
programmatic model of the information space and doesn't require any special
handling. It also emphasizes properties such as the closure of personalization
operators and the factorizability of information spaces, that help
relate design decisions to needs identified through scenario analysis.
While these same issues are pertinent in~\cite{schwabe2,schwabe1}, the
approach there is more reminiscent of design patterns (and integrating
requirements via OO analysis), whereas our idea is to use explanations to
identify opportunities for providing personalization facilities. PIPE's
approach also makes more effective use of domain-specific knowledge, both
embodied in the scenarios and assumed in the modeling of the
information seeking
activity.

As Russell and Norvig point out, empirical analysis of efficiency
is central to EBG~\cite{russel-norvig}. They emphasize 
that `the efficiency of an [information system factorization] is actually its 
average-case complexity
on a population of [scenarios that are likely to be encountered].' Being
too specific when operationalizing explanations will lead to making more
distinctions than losing them, contributing to lesser orthogonality (salience,
as used in~\cite{potts})
among scenarios. Defining operationality~\cite{keller-op} carefully
in the personalization context is an area for future research.

\section{Discussion}
\label{discuss}
This research makes contributions to the state-of-the-art in both
personalization systems and scenario-based design. For personalization, 
we have clarified the aspects of requirements specification and high-level 
elucidation of goals, showing how explanations from usage scenarios can provide 
models for PIPE. In particular, the problem of designing a representation has 
been formalized as a search through a space of PIPE models, driven
by the operationality criterion. The techniques presented in this paper can 
also be used to analyze existing personalization facilities, by determining if 
they address the requirements and opportunities of observed usage scenarios.

Our methodology of developing explanations of scenarios also adds value 
to the overall effort of creating scenario-based descriptions
of software and systems and is a further argument for adopting scenario-based
design (SBD) methods. By adopting the EBG view of operationalization, and 
for applications such as personalization, we can use a strong domain theory 
to reason about how scenarios can be effectively supported. In particular, our 
methodology helps to propositionalize information system designs.  

From the SBD viewpoint, PIPE emphasizes programmatic 
approaches of transforming between representations. We can extend our
approach to investigate other opportunities for partial evaluation and
also to include other program transformation techniques. 
This will support the provision of views on scenarios, performing
tree-manipulation operations, and propagating effects of changes through 
a representation. For instance, the paradigm of 
`training wheels in a user interface'~\cite{jack-wheels} which relies on 
masking functionality can be expressed using such methods. 

It should be remarked that both PIPE and the explanation-based view of
operationalization are two {\it specific} choices that we have made in 
understanding the early stages in the lifecycle of personalization
systems. Programs and partial evaluation serve the role of a
modeling methodology and transformation technique for 
information spaces; explanations supply the mechanism
that connects needs and requirements identified from SBD to
modeling choices in PIPE. While this is admittedly only one (and 
to our knowledge, the first) approach, it provides a glimpse into how 
other lifecycle models can be organized and how they will 
differ from models for general software systems. 

This investigation suggests many interesting avenues for future research.
It is good EBG tradition to identify novel ways in which explanations can
be generalized and personalization is fertile with opportunities.
For instance, PIPE models allow for out-of-turn 
interactions (by partial evaluation). This helps overcome the mismatch between
the user's mental model and the facilities available for describing the
information seeking goal. In a non-PIPE implementation, the user has to
manually reconcile this mismatch and perform an exploratory mode of seeking
before being able to use the system to satisfy the goal. We would
like to generalize such scenarios to recognize 
when `exploratory steps have been used because out-of-turn interaction was 
not possible.' For instance, this would allow us to explain that `Nancy
clicked on all links at the top-level page, not because this is what she
wanted but because she was exploring to see which one of them 
led to her choice of link at the second-level.' 
Another form of generalization pertains to the constructive induction of
intermediate subgoals in explanation structures. Recall that we employed
think-aloud protocols to augment our domain theory to support certain
explanations. Automating the induction of subgoals for recurring patterns
of interaction (such as manual information integration) is a possible
direction for future work.

The concept of operationality can be explored more carefully in the context
of personalization. Our study has exploited only two dimensions of 
operationality, namely the number of explanation trees and the operationality
boundaries in each. Once again EBG research~\cite{keller-op} suggests
other important dimensions --- such as variability, granularity, and
certainty --- which can be used to define an `operationality assessment
procedure.' Studying these concepts for information systems will allow 
the characterization of personalization applications in terms of 
operationality dimensions. For instance, differences between news-feed
customization services (e.g., {\tt myCNN.com}, PointCast) 
can be expressed in terms of operationality. Such characterizations will also 
aid 
in clarifying the concept of utility (and usability) of personalization systems, since
operationality is primarily concerned with empirical efficiency of models.

The notion of a personalization lifecycle can be usefully extended, to support
iterative improvement of PIPE models and to include stages like verification 
and validation. Support for iterative refinement is important in 
extending and composing existing personalization systems. It requires a tighter integration
between the explanation construction procedure and the way in which scenarios
are selected for explanation. Iterative improvement can also benefit from 
existing approaches to scenario repair and related EBG techniques such
as `learning by failing to explain~\cite{hall-fail}.' 
Methodologies for verification and validation can be incorporated 
in our framework, in the form of analytic and empirical frameworks for
utility analysis~\cite{dejong}. Such frameworks can take advantage of 
the characterization of the space of PIPE models produced by explaining 
scenarios and prior knowledge of the distribution of problem scenarios.

An emerging frontier involves modeling {\it
context} in information systems.
Consider:
\begin{descit}
\noindent
{\bf Person:} Remember the hotel where we hosted the annual convention?\\
{\bf Secretary:} Yes.\\
{\bf Person:} Reserve it for next Friday's event.
\end{descit}
Creating a personalization system that exploits context
amounts to storing
and retrieving smaller (or partial) explanations for use in constructing
larger-scope explanations.
Scenarios that do not permit complete explanations 
can also be interpreted as activities
for building and organizing context, for use in later situations. 
The explanation-based view of scenarios allows the decomposition of 
structures to aid in such reasoning.

Our systems-oriented view of personalization will find greater acceptance
if tools and software are available for automating various aspects of
the methodology. For instance, scenario management tools and explanation
engines can be prototyped for targeted information spaces. Specific
techniques for web mining and modeling user interactions can be incorporated
as reusable sub-explanations. This will allow us to design personalization
systems around existing system infrastructure. To aid in the maintainability
of PIPE models, specific scenario libraries (called `chunking~\cite{chunking}' in AI) and `frequently used explanations' can also be designed.

The central role played by the domain theory in our
methodology signifies a back-to-basics
approach in personalization system design. For a given information seeking
activity, a domain theory is characterized by its `explanatory power' and
how effectively it allows us to define the parameters of a personalization
space. This suggests that we should aim for a more fundamental understanding
of how domain theories characterize information spaces and the
situations to which they can be usefully applied. Our methodology is the 
first in which such questions can be directly expressed. Extending work
in these directions will help us to architect an information resource
for personalization and to provide rigorous metrics for evaluating the 
applicability of PIPE in a new situation. Together, they will take important 
steps in establishing a lifecycle of personalization system design. 

\section*{Acknowledgements}
Saverio Perugini helped implement the study
presented in the Appendix.
Marcos Gon\c{c}alves identified several pertinent references. Robert
Capra helped make connections from our work to
mixed-initiative interaction and contextual abstractions. All three colleagues
read drafts of this paper and provided important comments.

\bibliographystyle{plain}
\bibliography{final}

\newpage
\section{Appendix: An Example Application}
\label{case}
\begin{figure}
\centering
\begin{tabular}{cc}
\mbox{\psfig{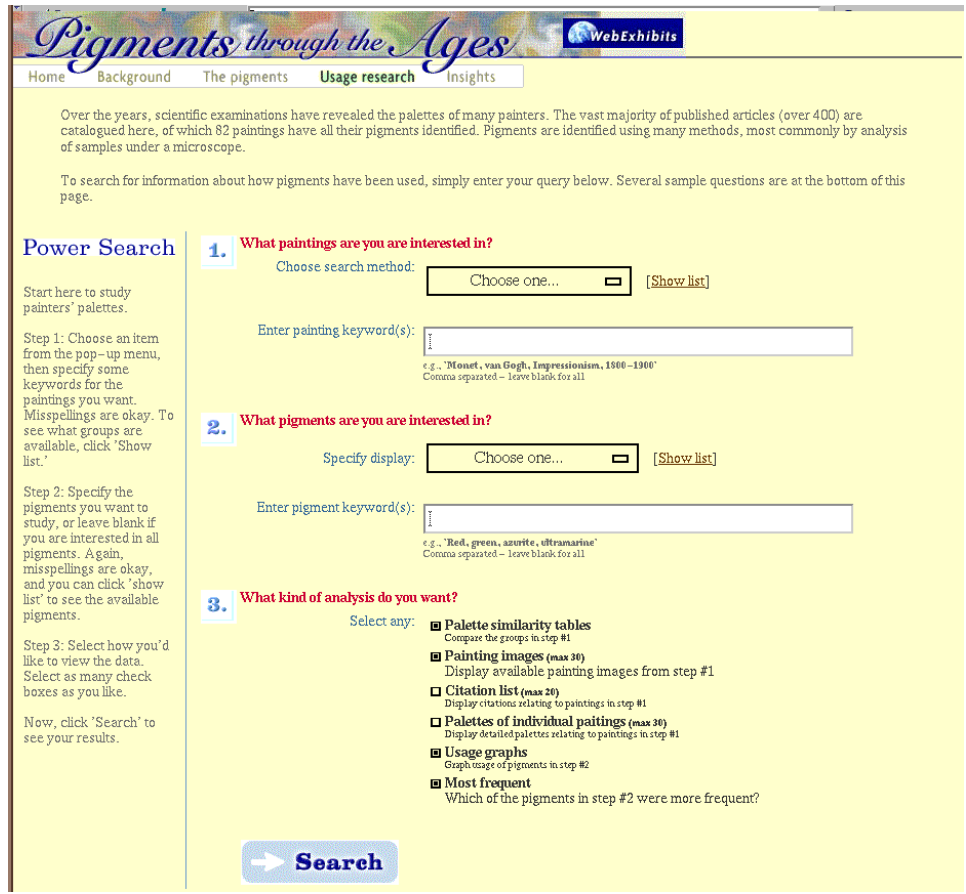}}
\end{tabular}
\caption{Original layout of the `Pigments through the Ages' website.}
\label{pigmentspicture}
\end{figure}

As a demonstrator of the ideas presented in this paper, we describe the
personalization of the `Pigments through the Ages' website at
[{\tt www.webexhibits.org/pigments}], a public service that uses
pigment analysis catalogs to identify and reveal the palettes of
painters in different eras and genres. As shown in the top-level interface
depicted in Fig.~\ref{pigmentspicture}, a variety of information resources
are modeled in this site. Users can search for paintings by artist, style,
period, or by membership in a particular pigment group. Notice also that the
interface in Fig.~\ref{pigmentspicture} provides some `hardwired' scenarios such
as comparing palette similarity tables or analyzing pigment usage in a certain
age. However, even a simple query such as `What is the influence of colors
from the baroque era on the neo-classic styles of paintings?'
cannot be accommodated without manual information integration because the
interaction sequences are hardwired. 

\begin{figure}
\centering
\framebox{
\begin{minipage}{2\colwidth}
\small
\begin{description}
\item[Jeremy's scenario]: Jeremy is attempting to compare how colors from
the baroque
era were used in the neo-classic paintings. In particular, he is interested
in usage graphs
for pigments in neo-classic that are most similar to ones used in baroque.
Jeremy surveys the existing facilities at
the site and chooses `by Artist, Style, or Period' as the {\tt search
method}. Next,
he specifies `neo-classic baroque' in the text box for {\tt painting
keywords.} He
(correctly) reasons that this specification identifies the paintings to be
used for analysis.
Next, he chooses `All pigments' from the {\tt specify display} dropbox and
selects
`Usage graphs' as the {\tt analysis kind}. He (incorrectly) assumes that
this will
compare every painting from baroque with every painting from neo-classic
and that the
system will present an usage
graph for each such comparison.
On inspecting the results Jeremy notices,
instead, that the usage graphs are for {\it all} pigments used in the set
\{neo-classic $\cup$ baroque\}, not quite
what he had in mind. He wonders for 5 minutes and realizes that the site
does not provide any direct
interface to specify his form of analysis.

Jeremy pursues an alternative strategy. He is going to first find the
common colors across
baroque and neo-classic, and then determine their usage patterns in
neo-classic.
He opens an additional browser window. In the new one,
he specifies `by Artist, Style, or Period' as the {\tt search method} and
`neo-classic baroque' as the {\tt painting keywords.} He clicks on the
palette similarity
table checkbox and obtains a matrix of values that indicate how colors from
one period were
utilized in another. As he expected, this time the specialized interface
interprets
that the two groups he specified in {\tt painting keywords} have to be
compared with each other.
The results page provides a matrix whose entries denote similarity levels.
He picks
out the pigments corresponding to the highest similarity levels and shifts
control to his old
browser window. There, he types in these pigments in the {\tt pigment
keywords} textbox and,
this time, types only `neo classic' in the {\tt painting keyword} textbox.
All other settings were
as he left them (including the `Usage graphs' request). This time, the
output screen provides a
histogram of the usage of the baroque pigments in neo-classic, which
satisfies Jeremy's
information seeking goal.
\end{description}
\end{minipage}}
\caption{A `Pigments through the Ages' scenario whose explanation was
subsequently operationalized.}
\label{exam-scenario}
\end{figure}

\vspace{-0.1in}
\subsubsection*{Problem Scenario Development}
A group of 10 participants were identified and instructed to explore the
layout and organization of information at this site. After a 
period of acquainting themselves with the site, they were asked to 
identify one specific query (or analysis) and use
the facilities at the site to answer their query. The exact interaction
sequences (including clicked hyperlinks, manual information integration) was 
recorded for all the participants. One such scenario is described in
Fig.~\ref{exam-scenario}.

\vspace{-0.1in}
\subsubsection*{Domain Theory}
The domain theory for this application was obtained from three sources.
The first was an explicit crawl of the site that outlined how interactions
result in specification of information seeking attributes. The second was
a `Background' webpage at [{\tt http://\hskip0ex webexhibits.\hskip0ex org/\hskip0ex pigments/\hskip0ex intro/\hskip0ex index.\hskip0ex html}]
that outlined a schema for how the website should
be used and how to browse through the various sections. For instance,
one mode of operation suggested at the site was to choose the `Usage Research'
category and see which pigments were used in different paintings. Another
mode of operation was to jump to a particular pigment page and then
browse through categories of information outlining technical details. 
All of these forms of navigation at the site were modeled as possibilities
for satisfying the top-level personalization goal. The third source
was from analyzing user interactions that revealed opportunities for
information integration across multiple pages of the site. Once again these
were modeled as specific possibilities of instantiating
the top-level personalization
goal. The domain theory was 
represented in CLIPS but only certain portions of the theory were materialized
when conducting explanations. We ensured that all 10 scenarios can
be explained by the domain theory.

\vspace{-0.1in}
\subsubsection*{Constructing and Analyzing Explanations} 
Explanations of user interactions revealed that starting from
either artists, paintings, or eras, the users systematically browsed
through subcategories or compared palettes to arrive at the relevant
pigments (used by that artist, in the painting, or in that era, 
respectively). Furthermore, all pigments share common modes of 
information seeking, such as browsing through their history of use, procedures
for preparation, and technical details of their chemical composition. 

\vspace{-0.1in}
\subsubsection*{Operationalization}
We hence operationalized the explanation structure(s) as two function
invocations in sequence, the first to determine an appropriate pigment
category, and the second to browse through the entries in that category by
various means. 
%The latter web source could be factored out and
%procedurally invoked at any point in the program where the name of the
%pigment has been resolved by other means. 
%
We thus arrived at a single
structure in support of all the 10 scenarios.
The factorization implied by the structure permits the following analysis:
\begin{descit}{}
For the pigment categories defined by $X$, provide the details involving $Y$.
\end{descit}
$X$ denotes information such as a genre, a style, a painting, or
a particular artist. $Y$ denotes features of pigments such as usage history,
chemical composition, and dyeing processes.
Each of $X$ and $Y$ could be defined either directly or
involving attributes of other entities that relate to them. For instance
$X$ could be a painting keyword such as `Rembrandt' (which means that we
are interested in pigments used by Rembrandt) or it could be the result of
the palette similarity function applied on two painting styles (which would
mean that we are interested in pigments that satisfy some acceptable 
threshold for similarity). In addition, there are dependencies among the 
allowed entries in $X$ and $Y$. 
%Since PIPE supports partial information, the above query
%can be invoked without any knowledge of $X$; this would just mean that
%a particular type of information is requested of all pigment categories
%(and hence, pigments).
%The operationalization implied by the above analysis indicated
%to us that we include ways to infer attributes of parent entities, given
%properties of sub-entities. For instance, the `bone black' category
%indicates that both the {\tt bone\_black} and {\tt black} program
%variables should be set. (This was possible only in structures that
%involved subtrees.)
Jeremy's scenario
satisfies the above template where
$X$ denotes the result of applying the palette similarity function to
`neo-classic baroque' and $Y$ denotes `usage graphs.'
%The above template for personalization
%is thus not currently supported by the facility, automatically.

\vspace{-0.1in}
\subsubsection*{Representation in PIPE}
To support the user in defining $X$ and $Y$, we modeled various information
sources such as
the catalog contents (which contains paintings from 950 to 1981),
the palette similarity table (which is just a function in our program),
citations of paintings, and auxiliary information such
as images, histories, where the painting is housed, and other legends.
Overlaps of painting styles across different periods contribute to sources 
of semistructure in this site and a corresponding reduction in the 
composite program size. Our composite
program was represented in 2369 lines of C code involving hundreds of program
variables that could be turned on or off with user input. Approaches
for modeling the various elements in this study are 
described in~\cite{pipe-tois}. We did not implement the mappings from the
(specialized) program back to the information space because we only wanted
to evaluate the effectiveness of our modeling (more on this below).

\vspace{-0.1in}
\subsubsection*{Evaluation}
The evaluation of systems designed with PIPE is an interesting issue
in itself; we address this topic in greater detail in~\cite{pipe-tois}.
While user satisfaction surveys show convincing results (see, for
instance~\cite{naren-ic}), PIPE is more a modeling methodology for
personalization, and not a system per se. As such, its effectiveness
depends on what is modeled (and how). The research presented in this
paper gives us a direct way to assess the modeling capability of PIPE.

We identified a test group of 15 users (different from those
who participated in the original scenario analysis)
and asked them to experiment with
the unpersonalized pigments site. Each of them
was then asked to identify and carefully describe 2-3 personalization
scenarios. In total, 35 scenarios were identified.
An example is the following analysis:
\begin{descit}{}
What are the symbolic connotations of pigments used by artists in the
Renaissance era?
\end{descit}
(One of the answers to this query is a web page that describes the
interpretation of red as invincibility
in Jan van Eyck's 1434 classic {\it Arnolfini Wedding}.)
We then evaluated our PIPE representation by the fraction
of scenarios that can be described in our modeling (and are hence
amenable to personalization by partial evaluation). 
All scenarios
except two passed our test. The two unmodelable scenarios involved
the `Orpiment' pigment which was listed in both
the `Yellow' and `Orange' categories and was variously referred to by users
as belonging to one, but not the other.
This ambiguity implies that our modeling did not contain
sufficient information to complete the proof (i.e., it
could not uniquely distinguish between these two distinct specifications
involving $X$). More contextual information needs to be encoded in our
modeling so that this ambiguity is resolved. 

A full listing of the scenarios used
in evaluation follows. Except for scenarios~\ref{orp1} and ~\ref{orp2}, 
all others can be
supported. Scenarios~\ref{first1},~\ref{orp2},~\ref{pref2},
~\ref{pref3}, and~\ref{pref4} indicate preferences for presentation which
can be addressed
when we recreate the personalized pages from the specialized program.
Scenario~\ref{strange} states preferences
for interactions at many levels. 
This amounts to repeated partial
evaluations of the information
space, in the order of attributes stated by that user.
Scenarios such as 27 imply a
desire to use complete evaluation with the designed PIPE model, not
partial evaluation. 
\begin{enumerate}
\item 
\label{first1}
What are the symbolic connotations of pigments used by artists in the
Renaissance era? Arrange the results on a single page,
in alphabetical order of pigments.
\item (similar to Jeremy's scenario) I would like to see how colors used in
1800-1900 have influenced paintings in the early part of the 20th
century. Show usage graphs for pigments that are similar across these
eras.
\item I would like to specify a pigment choice, not based on
a property of
painting, style, or era. Rather, I want all pigments for which descriptions of
chemical composition are available. So, if a pigment does not have this
information,
it should not be listed.
\item I would like to browse through pigment details at the root page, not
go through
information about paintings or painters.
\item \label{orp1} I
am interested to see how usage of pigments of a subcategory compares with that
of pigments in the parent category. For instance, does Orpiment usage
correlate with
usage of Yellow in 1800-1900 paintings?
\item The site facility lists only at most 10 palettes at a time. If I need
more, I have to
carefully pose multiple subqueries so that each of them does not involve
more than
10 paintings. Can you fix this problem so that I can see all palettes?
\item My period specifications don't seem to work at the site. When I manually
browse the site, I see annotations such as `1900-2000' and `1650-1750.' But
when
I pose my range as `1875-1925,' the site doesn't seem to understand.
Should I have to break my range up into these prespecified ranges? That seems
cumbersome.
\item \label{orp2}
Can I get a listing of pigments used by the Impressionists
along with their parent categories, side-by-side on a single page?
\item I would like the pigments arranged by history (e.g., middle ages),
followed
by an organization along countries.
\item I would like to be able to
choose a pigment according to ease of preparation.
\item I would like to search for pigments using a combination of
two criteria (such as geographical use and time period), but the site
allows only
one at a time.
\item I would like the interaction to proceed as follows: At the first level,
I will make a choice of history of usage, after which I will browse the site in
the traditional manner.
\item \label{pref2}
I would like pigments of the Blue category to be identified in alphabetical
order.
Then I would like to see the swatches of paints from the top five to be placed
alongside swatches of paint from the bottom five. This will show me the
range of intensities
of Blue available.
\item I would like a listing of pigments by their chemical name, not their
colloquial names.
\item I find myself repeatedly browsing pigments' pages to study the
fascinating stories of
how these pigments originate. Some of these pages don't seem to have any
stories. Can you
provide me a listing of only those pigments that have stories?
\item Which are the pigments that have German names or equivalents?
\item I would like to see the descriptions of pigments that have pictures in
them. I am not interested in purely textual descriptions.
\item I would like to directly select a specific pigment from a list, on the
first page.
\item Which pigments have been used by Alchemists?
\item \label{pref3}
Can you cascade the brief descriptions of pigments and remove all the other
information
pertaining to preparation and technical details?
\item Which pigments have citations to them? I would like a listing of only
those.
\item \label{pref4}
I am interested in the Green earth pigment.
I would like a page that has pictures of paintings and along with each, a
picture of the
swatch of paint from Green earth. This is just so that it is visually easy
to see how
much the painting emphasizes Green earth.
\item I would like pigments arranged by the year in which they were first
introduced.
\item I am interested in making pigments.
Can you please instruct me how to make every
pigment in the purple category?
\item \label{strange} I am interested in the citation lists for green pigments.
However, I would like to browse them by first making a selection of artist.
Then I will
select a period. And finally I will select among titles, if there are
choices. For the green pigments
used in these titles, I would like to see the citations.
\item How is the name for the Azurite pigment derived? What is the word origin?
\item Produce the 3D model for Titanium dioxide.
\item I know that Kandinsky has suggested that black indicates an inner
harmony of silence.
Can I see which forms of blacks were used in his paintings?
\item Can you give information about how pigments are used for body art,
tattoos
and other non-conventional forms of paintings?
\item What are the time periods when Lemon Yellow was used?
\item For paintings by Monet, can you display the top five most frequent
pigments?
\item What forms of white pigments have been used in paintings? Which ages were
they introduced in?
\item Give a histogram of how chrome yellow has been used over the times.
\item Arrange histograms of all purple pigments used after the 17th century.
\item How do pigments used by Picasso compare in usage with pigments used
in the 1920s,
in general?
\end{enumerate}
\end{document}